\documentclass[reprint,superscriptaddress,amsmath,longbibliography,amssymb,prl]{revtex4-1}

\usepackage{graphicx}% Include figure files
\usepackage{dcolumn}% Align table columns on decimal point
\usepackage{bm}% bold math
\usepackage{multirow}
\usepackage{array}
\usepackage{braket}
\usepackage[usenames,dvipsnames]{xcolor}
\usepackage{mathrsfs}
\usepackage{amsbsy}
\definecolor{nblue}{rgb}{0.0, 0.0, 1.0}
\definecolor{magenta}{rgb}{0.79, 0.08, 0.48}
\usepackage[colorlinks,linkcolor=nblue,urlcolor=magenta,citecolor=magenta,plainpages=false,pdfpagelabels,breaklinks]{hyperref}

\newcommand{\beq}{\begin{equation}}
\newcommand{\eeq}{\end{equation}}
\newcommand{\bea}{\begin{eqnarray}}
\newcommand{\eea}{\end{eqnarray}}

\newcommand{\bk} { \bm{k} }
\newcommand{\bp} { \bm{p} }
\newcommand{\bd} { \bm{d} }
\newcommand{\eqn}[1] {Eqn.~(\ref{#1})}
\newcommand{\fig}[1]{Fig.~\ref{#1}}

\newcommand{\mylabel}[1]{\label{#1}}

\begin{document}

%\preprint{}

\title{Chiral singlet superconductivity in the weakly correlated metal LaPt$_{3}$P}
%\title{Chiral singlet superconductivity in the centrosymmatric pnictide compound LaPt$_{3}$P}
%\title{Chiral $d$-wave superconductivity in LaPt$_{3}$P}
%\title{Nodal gap structure and time-reversal symmetry breaking in the unconventional superconductor LaPt$_{3}$P}
%\title{Time-reversal symmetry breaking in the nodal superconductor LaPt$_{3}$P}



%\author{}
%\thanks{}
%\affiliation{}

\author{P.\ K.\ Biswas}
\email{pabitra.biswas@stfc.ac.uk}
\affiliation{ISIS Pulsed Neutron and Muon Source, STFC Rutherford Appleton Laboratory, Harwell Campus, Didcot, Oxfordshire OX11 0QX, United Kingdom}

\author{S.\ K.\ Ghosh}
\email{S.Ghosh@kent.ac.uk}
\affiliation{School of Physical Sciences, University of Kent, Canterbury CT2 7NH, United Kingdom}

\author{J.\ Z.\ Zhao}
\affiliation{Co-Innovation Center for New Energetic Materials, Southwest University of Science and Technology, Mianyang, 621010, China} 

\author{D.\ A.\ Mayoh}
\affiliation{Physics Department, University of Warwick, Coventry, CV4 7AL, United Kingdom} 

\author{N.\ D.\ Zhigadlo}
\affiliation{Laboratory for Solid State Physics, ETH Zurich, 8093 Zurich, Switzerland}
\affiliation{CrystMat Company, 8037 Zurich, Switzerland} 

\author{Xiaofeng Xu}
\affiliation{Department of Applied Physics, Zhejiang University of Technology, Hangzhou 310023,China} 

\author{C.\ Baines}
\affiliation{Laboratory for Muon Spin Spectroscopy, Paul Scherrer Institute, CH-5232 Villigen PSI, Switzerland} 

\author{A.\ D.\ Hillier}
\affiliation{ISIS Pulsed Neutron and Muon Source, STFC Rutherford Appleton Laboratory, Harwell Campus, Didcot, Oxfordshire OX11 0QX, United Kingdom} 

\author{G.\ Balakrishnan}
\affiliation{Physics Department, University of Warwick, Coventry, CV4 7AL, United Kingdom} 

\author{M.\ R.\ Lees}
\affiliation{Physics Department, University of Warwick, Coventry, CV4 7AL, United Kingdom} 


\date{\today}
\maketitle

%\begin{abstract}
\textbf{
Topological superconductors (SCs) are novel phases of matter with nontrivial bulk topology. They host at their boundaries and vortex cores zero-energy Majorana bound states, potentially useful in fault-tolerant quantum computation~\cite{Sato2017}. Chiral SCs~\cite{Kallin2016} are particular examples of topological SCs with finite angular momentum Cooper pairs circulating around a unique chiral axis, thus spontaneously breaking time-reversal symmetry (TRS). They are rather scarce and usually feature triplet pairing: best studied examples in bulk materials are UPt$_3$ and Sr$_2$RuO$_4$ proposed to be $f$-wave and $p$-wave SCs respectively, although many open questions still remain~\cite{Kallin2016}. Chiral triplet SCs are, however, topologically fragile with the gapless Majorana modes weakly protected against symmetry preserving perturbations in contrast to chiral singlet SCs~\cite{Kobayashi2014,Goswami2015}. Using muon spin relaxation ($\mu$SR) measurements, here we report that the weakly correlated pnictide compound LaPt$_3$P has the two key features of a chiral SC: spontaneous magnetic fields inside the superconducting state indicating broken TRS and low temperature linear behaviour in the superfluid density indicating line nodes in the order parameter. Using symmetry analysis, first principles band structure calculation and mean-field theory, we unambiguously establish that the superconducting ground state of LaPt$_3$P is chiral $d$-wave singlet.}
%\end{abstract}
%\maketitle
 
Cooper pairs in conventional SCs, such as the elemental metals, form due to pairing of electrons by phonon-mediated attractive interaction into the most symmetric $s$-wave spin-singlet state~\cite{Tinkham}. In contrast, unconventional SCs defined as having zero average onsite pairing amplitude pose a pivotal challenge in resolving how superconductivity emerges from a complex normal state. They usually require a long-range interaction~\cite{Scalapino2012} and have lower symmetry Cooper pairs. A special class of unconventional SCs are the chiral SCs. A well established realization of a chiral $p$-wave triplet state is the $A$-phase of superfluid He$^3$~\cite{Schnyder2015}. In addition to UPt$_3$ and Sr$_2$RuO$_4$, the heavy fermion SC UTe$_2$ is also proposed to be a chiral triplet SC~\cite{Jiao2020}. The chiral singlet SCs are, however, extremely rare and are proposed to be realized within the hidden order phase of the strongly correlated heavy fermion SC URu$_2$Si$_2$~\cite{Mydosh2011} and in the locally noncentrosymmetric material SrPtAs~\cite{Biswas2013} with many unresolved issues.

%Muon spin rotation and relaxation ($\mu$SR) technique is a very sensitive probe of local magnetic field distribution within a material and is in high demand to explore the pairing symmetry and time-reversal symmetry (TRS) breaking in USCs~\cite{Sonier2000}. Until quite recently, TRS was known to be broken only in a limited number of USCs~\cite{Ghosh2020a}, examples include the filled-skutterudite PrOs$_4$Sb$_{12}$, Sr$_2$RuO$_4$ and uranium based heavy fermions UPt$_3$ and (U, Th)Be$_{13}$. $\mu$SR played a central role in the recent discoveries~\cite{Ghosh2020a} of a growing number of noncentrosymmetric (materials lacking inversion symmetry) superconductors such as LaNiC$_2$, Re$_6$(Zr, Hf, Ti), La$_7$(Ir, Rh)$_3$ and Zr$_3$Ir, and a few centrosymmetric superconductors such as LaNiGa$_2$ and (Lu, Y, Sc)Rh$_6$Sn$_{18}$ breaking TRS below T$_c$. Most of these materials show fully gapped excitation spectrum insensitive to nonmagnetic impurities and the role of inversion symmetry in determining the phenomenon of TRS breaking is not clear. It is thus imperative to find an USC which has a nodal gap structure and also show TRS breaking. 

LaPt$_{3}$P is a member of the platinum pnictide family of SCs $A$Pt$_3$P ($A$ = Ca, Sr and La) with a centrosymmetric primitive tetragonal structure~\cite{Takayama2012}. Its $T_{\rm c} = 1.1$ K is significantly lower than its other two isostructural counterparts SrPt$_{3}$P ($T_{\rm c} = 8.4$ K) and CaPt$_{3}$P ($T_{\rm c} = 6.6$ K)~\cite{Takayama2012} which are conventional Bardeen-Cooper-Schrieffer (BCS) SCs. Indications of the unconventional nature of superconductivity in LaPt$_{3}$P come both from theory: first principles Migdal-Eliashberg-theory~\cite{Subedi2013} and experiments: very low $T_{\rm c}$, unsaturated resistivity up to room temperature and a weak specific heat jump at $T_{\rm c}$~\cite{Takayama2012}. The chiral nature of superconductivity of LaPt$_{3}$P with topologically protected Majorana Fermi-arc and Majorana flat-band, which we uncover here, fits nicely with these characteristics.\\% The proposed chiral $d$-wave state of LaPt$_{3}$P has two types of topologically protected nodes: two Weyl nodes at the two poles and a equatorial line node which dominates the low temperature thermodynamics properties due to larger low energy density of states (DOS). They are expected to have characteristic transport signatures such as zero-bias conductance peaks and anomalous thermal Hall effect~\cite{Schnyder2015,Goswami2013,Goswami2015}.\\ 

\noindent\textbf{Experimental results}\\
We have performed a comprehensive analysis of the superconducting properties of LaPt$_{3}$P using the $\mu$SR technique. Two sets of polycrystalline LaPt$_{3}$P specimens, referred to here as sample-A (from Warwick, UK) and sample-B (from ETH, Switzerland), were synthesized at two different laboratories by completely different methods. Zero-field (ZF), longitudinal-field (LF), and transverse-field (TF) $\mu$SR measurements were performed on these samples at two different muon facilities: sample-A in the MUSR spectrometer at the ISIS Pulsed Neutron and Muon Source, UK, and sample-B in the LTF spectrometer at the Paul Scherrer Institut (PSI), Switzerland.

\begin{figure}[ht]
\begin{center}
\includegraphics[width=0.9\columnwidth]{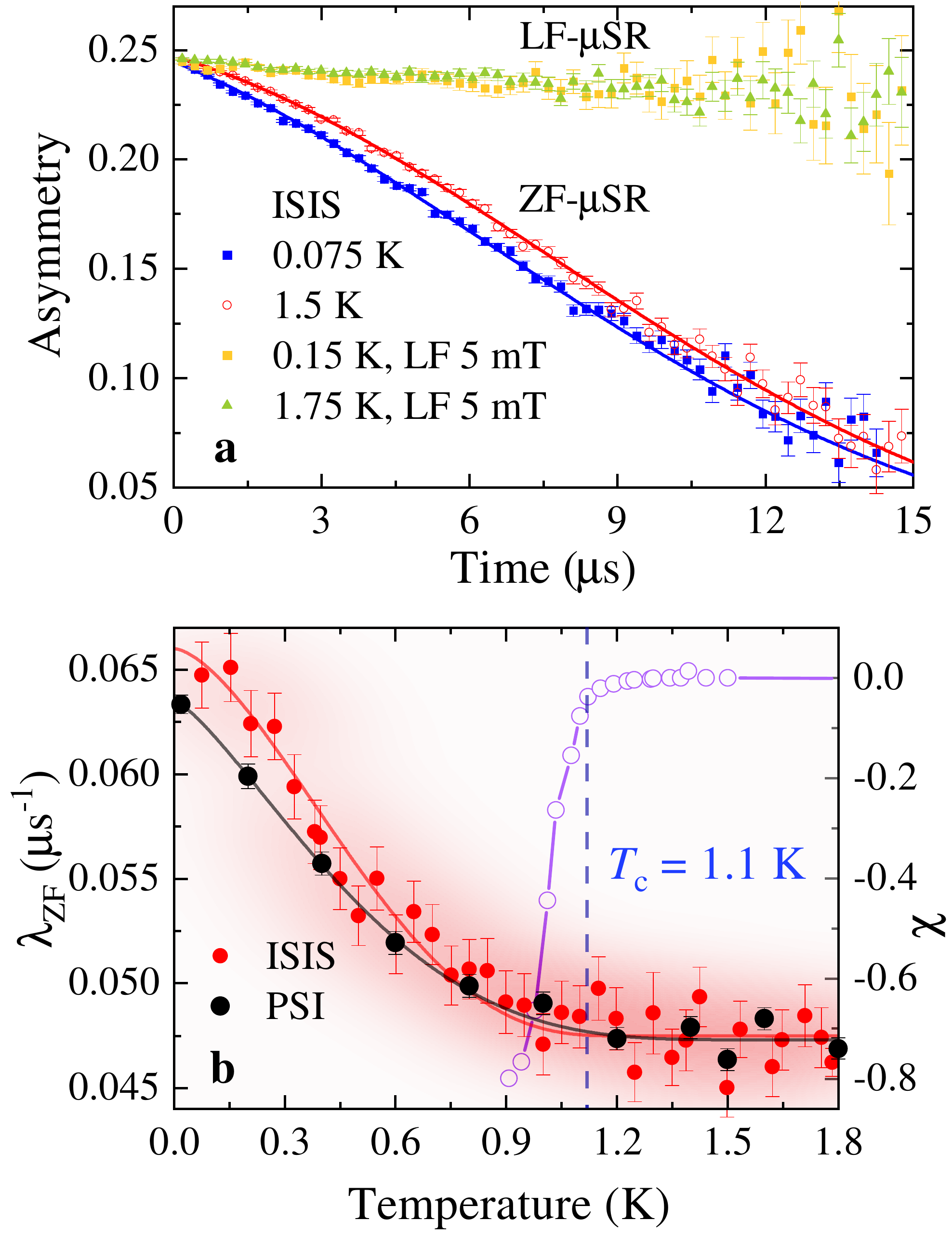}
\vspace{-0.25cm}
%\caption{(Color online)
\caption{\textbf{Evidence of TRS-breaking superconductivity in LaPt$_{3}$P by ZF-$\mu$SR measurements.} \textbf{a}) ZF-$\mu$SR time spectra collected at 75~mK and 1.5~K for sample-A of LaPt$_{3}$P. The solid lines are the fits to the data using Eq.~\ref{eq:KT_ZFequation}. \textbf{b}) The temperature dependence of the extracted $\lambda_{\rm ZF}$ (left axis) for sample-A (ISIS) and sample-B (PSI) showing a clear increase in the muon spin relaxation rate below $T_{\rm c}$. The PSI data has been shifted by $0.004$ $\mu$s$^{-1}$ to match the baseline value of the ISIS data. Variation of the zero-field-cooled magnetic susceptibility ($\chi$) on the right axis for sample-B.}
 \label{fig:ZF}
\end{center}
\end{figure}

ZF-$\mu$SR measurements reveal spontaneous magnetic fields arising just below $T_{\rm c} \approx 1.1$ K (example charcterization is shown by the zero-field-cooled magnetic susceptibility ($\chi$) data for sample-B on the right axis of \fig{fig:ZF}\textbf{b}) associated with a TRS breaking superconducting state in both samples of LaPt$_{3}$P, performed on different instruments. \fig{fig:ZF}\textbf{a} shows representative ZF-$\mu$SR time spectra of LaPt$_{3}$P collected at $75$~mK (superconducting state) and at $1.5$~K (normal state) on sample-A at ISIS. The data below $T_{\rm c}$ show a clear increase in muon-spin relaxation rate compared to the data collected in the normal state. To unravel the origin of the spontaneous magnetism at low temperature, we collected ZF-$\mu$SR time spectra over a range of temperatures across $T_{\rm c}$ and extracted temperature dependence of the muon-spin relaxation rate by fitting the data with a Gaussian Kubo-Toyabe relaxation function $\mathcal{G}(t)$~\cite{yaouanc2011} multiplied by an exponential decay:
\beq\mylabel{eq:KT_ZFequation}
A(t)= A(0) \mathcal{G}(t) {\exp}(-\lambda_{\rm ZF} t) + A_{\rm bg}
\eeq
where, $A(0)$ and $A_{\rm bg}$ are the initial and background asymmetries of the ZF-$\mu$SR time spectra, respectively. $\mathcal{G}(t) = \frac{1}{3}+\frac{2}{3}\left(1-\sigma_{\rm ZF}^2t^2\right){\exp}\left(-\sigma_{\rm ZF}^2t^2/2\right)$. $\sigma_{\rm ZF}$ and $\lambda_{\rm ZF}$ represent the muon spin relaxation rates originating from the presence of nuclear and electronic moments in the sample, respectively. In the fitting, $\sigma_{\rm ZF}$ is found to be nearly temperature independent and hence fixed to the average value of $0.071(4)$ $\mu$s$^{-1}$ for sample-A and $0.050(3)$ $\mu$s$^{-1}$ for sample-B. The temperature dependence of $\lambda_{\rm ZF}$ is shown in Fig.~\ref{fig:ZF}\textbf{b}. $\lambda_{\rm ZF}$ has a distinct systematic increase  below $T_{\rm c}$ for both the samples which implies that the effect is sample and spectrometer independent. Moreover, the effect can be suppressed very easily by a weak longitudinal field of $5$ mT for both the samples. It is shown in Fig.~\ref{fig:ZF}\textbf{a} for sample-A. This strongly suggests that the additional relaxation below $T_c$ is not due to rapidly fluctuating fields~\cite{Hayano1980}, but rather associated with very weak fields which are static or quasistatic on the time-scale of muon life-time. The spontaneous static magnetic field arising just below $T_{\rm c}$ is so intimately connected with superconductivity that we can safely say its existence is direct evidence for TRS-breaking superconducting state in LaPt$_{3}$P. From the change $\Delta\lambda_{\rm ZF} = \lambda_{\rm ZF}(T \approx 0) - \lambda_{\rm ZF}(T>T_c)$ we can estimate the corresponding spontaneous internal magnetic field at the muon site $B_{\rm int} \approx \Delta\lambda_{\rm ZF}/\gamma_\mu=0.22(4)$ G for sample-A and $0.18(2)$ G for sample-B which are very similar to that of other TRS breaking SCs~\cite{Ghosh2020}. Here, $\gamma_{\mu}/(2\pi)=13.55$~kHz/G is the muon gyromagnetic ratio. 

\begin{figure}[htb]
\includegraphics[width=0.95\columnwidth]{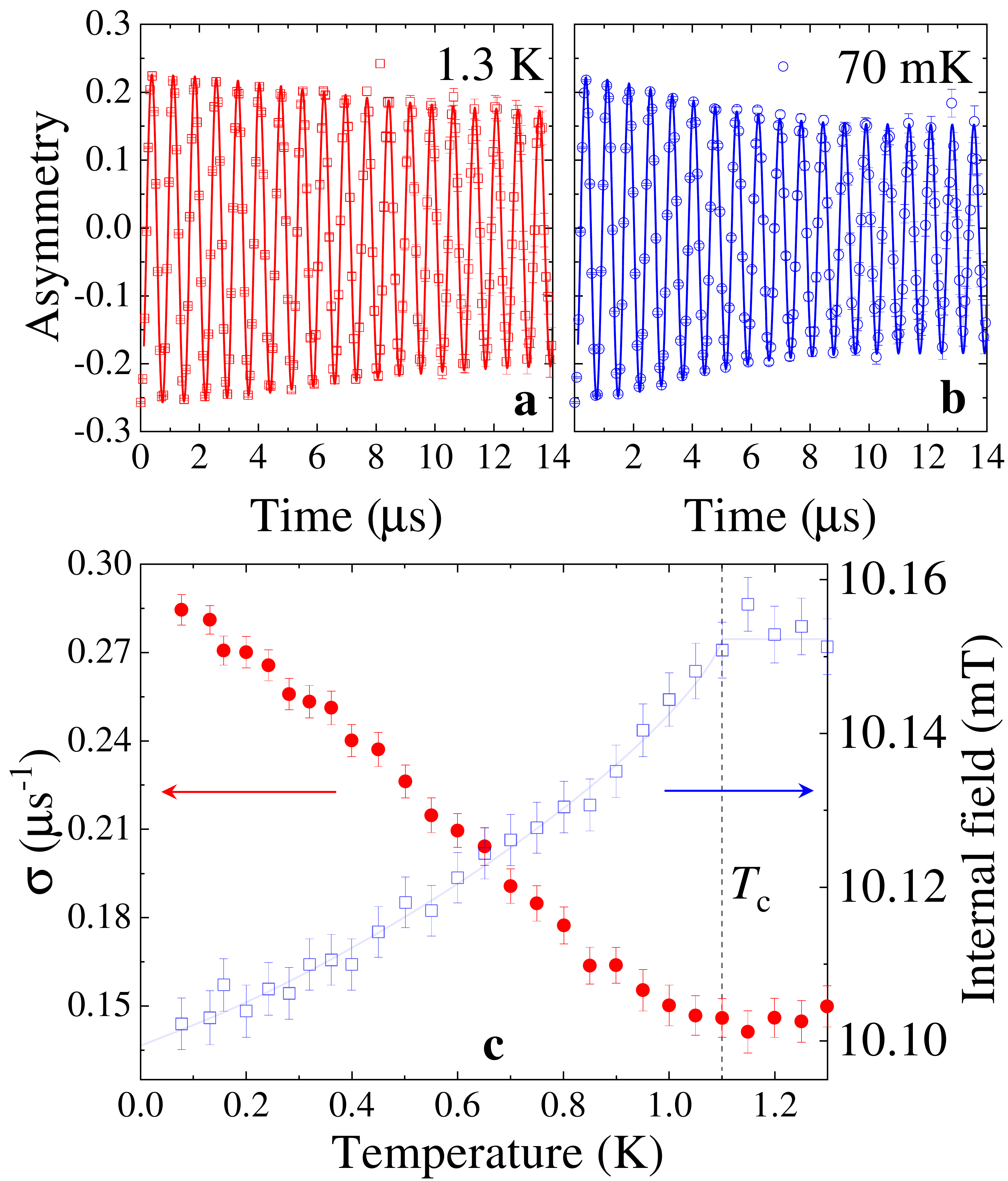}
\vspace{-0.25cm}
%\caption{(Color online)
\caption{ \textbf{Superconducting properties of LaPt$_{3}$P by TF-$\mu$SR measurements.} TF-$\mu$SR time spectra of LaPt$_{3}$P collected at \textbf{a}) 1.3~K and \textbf{b}) 70~mK for sample-A in a transverse field of 10~mT. The solid lines are the fits to the data using Eq.~\ref{Depolarization_Fit}. \textbf{c}) The temperature dependence of the extracted $\sigma$ (left panel) and internal field (right panel) of sample-A.}
 \label{fig:TF}
\end{figure}

\begin{figure}[htb]
\includegraphics[width=0.95\columnwidth]{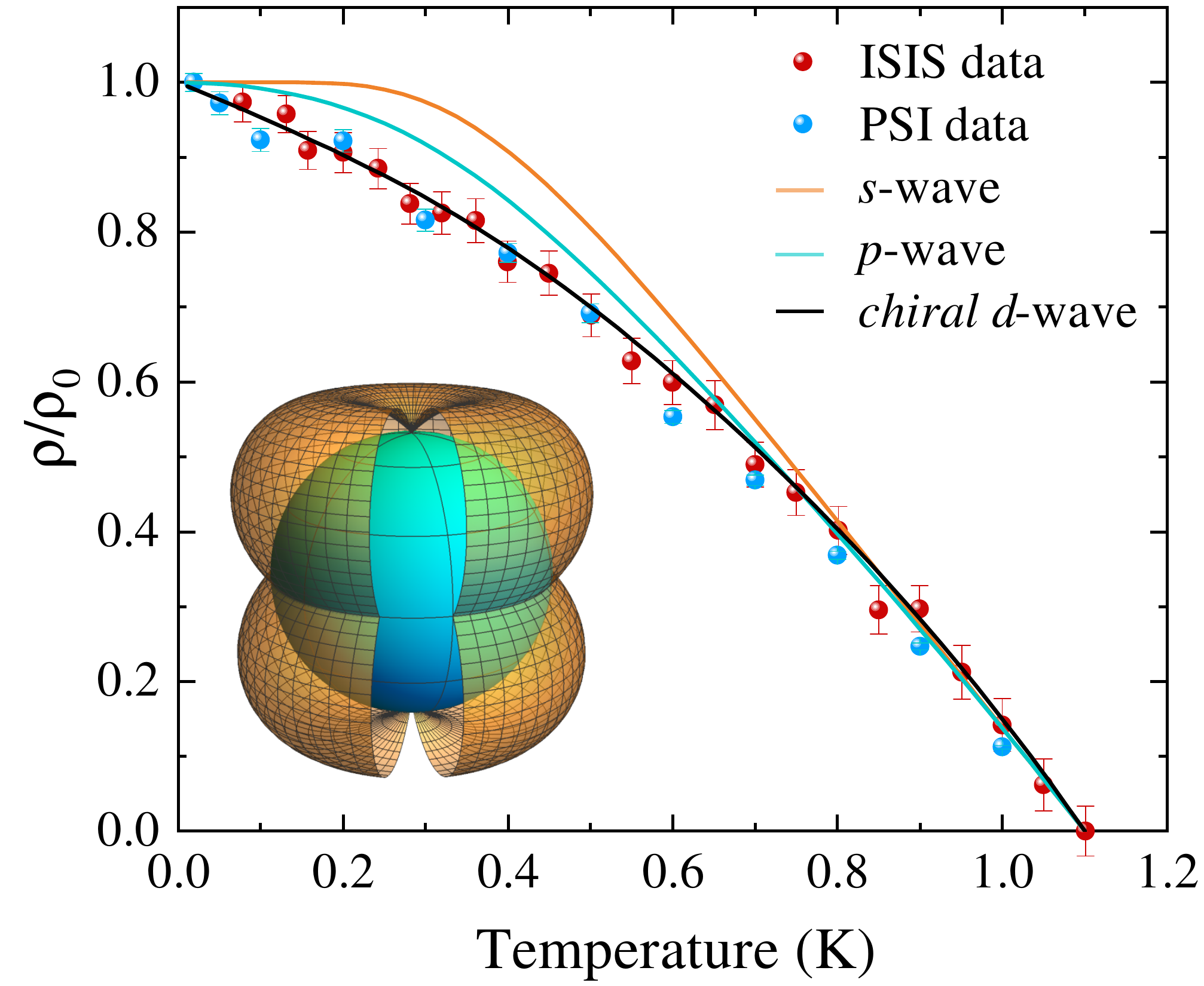}
\vspace{-0.25cm}
%\caption{(Color online)
\caption{ \textbf{Evidence of \textit{chiral $d$-wave} superconductivity in  LaPt$_{3}$P.} Superfluid density ($\rho$) of LaPt$_{3}$P as a function of temperature normalized by its zero-temperature value $\rho_0$. The solid lines are fits to the data using different models of gap symmetry. Inset shows the schematic representation of the nodes of the \textit{chiral $d$-wave} state.}
 \label{fig:sfdensity}
\end{figure}

\begin{figure*}[htb]
\includegraphics[width=\linewidth]{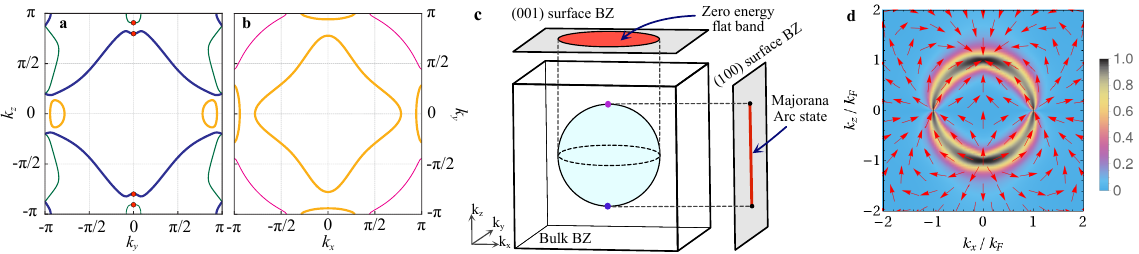}
\vspace{-0.5cm}
%\caption{(Color online)
\caption{ \textbf{Properties of the normal and superconducting states of LaPt$_3$P.} Projections of the four Fermi surfaces of LaPt$_3$P with SOC on the $y-z$ plane in \textbf{a} and $x-y$ plane in \textbf{b}. The thickness of the lines are proportional to the contribution of the Fermi surfaces to the DOS at the Fermi level (green-- $10.3\%$, blue-- $43.4\%$, orange-- $40\%$ and magenta-- $6.3\%$). The point nodes of the \textit{chiral $d$-wave} gap are shown by red dots in \textbf{a} and the line node reside on the $x-y$ plane in \textbf{b}. \textbf{c}) Schematic view of the Majorana Fermi-arc and the zero energy Majorana flat-band corresponding to the two Weyl point nodes and the line node respectively on the respective surface Brillouin zones (BZs) assuming a spherical Fermi surface. \textbf{d}) Berry curvature $\mathbf{F}(\bk)$ corresponding to the two Weyl nodes on the $x-z$ plane. Arrows show the direction of $\mathbf{F}(\bk)$ and the colour scale shows its magnitude $=\frac{2}{\pi}\arctan(|\mathbf{F}(\bk)|)$. $\Delta_0 = 0.5 \mu$ was chosen for clarity while a more realistic weak-coupling limit $\Delta_0 \ll \mu$ gives a more sharply peaked curvature at the Fermi surface.}
\label{fig:SurfaceStates}
\end{figure*}

We show the TF-$\mu$SR time spectra for sample-A in \fig{fig:TF}\textbf{a} and \fig{fig:TF}\textbf{b} at two different temperatures. The spectrum in \fig{fig:TF}\textbf{a} shows only weak relaxation mainly due to the transverse (2/3) component of the weak nuclear moments present in the material in the normal state at $1.3$~K. In contrast, the spectrum in \fig{fig:TF}\textbf{b} in the superconducting state at $70$~mK shows higher relaxation due to the additional inhomogeneous field distribution of the vortex lattice, formed in the superconducting mixed state of LaPt$_{3}$P. The spectra are analyzed using the Gaussian damped spin precession function \cite{yaouanc2011}:
\begin{multline}
\label{Depolarization_Fit}
A_{TF}(t)=A(0)\exp\left(-\sigma^{2}t^{2}\right/2)\cos\left(\gamma_\mu \left\langle B\right\rangle t +\phi\right)\\
+A_{\rm bg}\cos\left(\gamma_\mu B_{\rm bg}t +\phi\right).
\end{multline}
Here $A(0)$ and $A_{\rm bg}$ are the initial asymmetries of the muons hitting and missing the sample respectively. $\left\langle B\right\rangle$ and $B_{\rm bg}$ are the internal and background magnetic fields, respectively. $\phi$ is the initial phase and $\sigma$ is the Gaussian muon spin relaxation rate of the muon precession signal. The background signal is due to the muons implanted on the outer silver mask where the relaxation rate of the muon precession signal is negligible due to very weak nuclear moments in silver. \fig{fig:TF}\textbf{c} shows the temperature dependence of $\sigma$ and internal field of sample-A. $\sigma (T)$ shows a change in slope at $T = T_{\rm c}$ which keeps on increasing with further lowering of temperature. Such an increase in $\sigma (T)$ just below $T_{\rm c}$ indicates that the sample is in the superconducting mixed state and the formation of vortex lattice has created an inhomogeneous field distribution at the muon sites. The internal fields felt by the muons show a diamagnetic shift in the superconducting state of LaPt$_{3}$P, a clear signature of bulk superconductivity in this material.

 The true contribution of the vortex lattice field distribution to the relaxation rate $\sigma_{\rm sc}$ can be estimated as $\sigma_{\rm sc} = (\sigma^2 - \sigma^2_{\rm nm})^{1/2}$, where $\sigma_{\rm nm}=0.1459(4)$~$\mu$s$^{-1}$ is the nuclear magnetic dipolar contribution assumed to be temperature independent. Within the Ginzburg-Landau theory of the vortex state, $\sigma_{\rm sc}$ is related to the London penetration depth $\lambda$ of a SC with high upper critical field by the Brandt equation~\cite{Brandt2003}:
\beq
\frac{\sigma_{sc}\left(T\right)}{\gamma_\mu}=0.06091\frac{\Phi_0}{\lambda^{2}\left(T\right)},
\eeq
where $\Phi_0=2.068\times10^{-15}$~Wb is the flux quantum. The superfluid density $\rho \propto \lambda^{-2}$. \fig{fig:sfdensity} shows the temperature dependence of $\rho$ for LaPt$_{3}$P. It clearly varies with temperature down to the lowest temperature $70$~mK and shows a linear increase below ${T_{\rm c}}/3$. This nonconstant low temperature behaviour is a signature of nodes in the superconducting gap. 

The pairing symmetry of LaPt$_{3}$P can be understood by analysing the superfluid density data using different models of the gap function $\Delta_{\bk}(T)$. For a given pairing model, we compute the superfluid density ($\rho$) as
\beq
\rho = 1 + 2 \bigg< \int_{\Delta_{\bk}(T)}^{\infty} \frac{E}{\sqrt{E^2 - |\Delta_{\bk}(T)|^2}} \frac{\partial f}{\partial E} dE \bigg>_{\text{FS}}.
\eeq
Here, $f=1/\left(1+e^{\frac{E}{k_B T}}\right)$ is the Fermi function and $\langle \rangle_{\text{FS}}$ represents an average over the Fermi surface (assumed to be spherical). We take $\Delta_{\bk}(T) = \Delta_m(T)g(\bk)$ where we assume a universal temperature dependence $\Delta_m(T) = \Delta_m(0)\tanh\left[1.82\left\{1.018\left(T_{\rm c}/T-1\right)\right\}^{0.51}\right]$~\cite{Carrington2003} and the function $g(\bk)$ contains its angular dependence. We use three different pairing models: $s$-wave (single uniform superconducting gap), $p$-wave (two point nodes at the two poles) and chiral $d$-wave (two point nodes at the two poles and a line node at the equator as shown in the inset of \fig{fig:sfdensity}). The fitting parameters are given in the Supplemental Material. We note from \fig{fig:sfdensity} that both the $s$-wave and the $p$-wave models lead to saturation in $\rho$ at low temperatures which is clearly not the case for LaPt$_{3}$P and the chiral $d$-wave model gives an excellent fit down to the lowest temperature. Nodal SCs are rare since the SC can gain condensation energy by eliminating nodes in the gap. Thus the simultaneous observation of nodal and TRS-breaking superconductivity makes LaPt$_3$P a unique material.  
\\

% Indeed, the $d$-wave model fit gives a much lower $\chi_{\rm reduced}^2$ value of 1.14 than the $s$-wave model (2.7) thus strongly suggesting the presence of nodes in the gap structure. For the $d$-wave model, the fit yields $\lambda(0)=650(10)$~nm and $\frac{\Delta(0)}{k_B T_c} = 2.48(2)$ which is larger than the weak coupling BCS value of 1.76 implying strong-coupling superconductivity in LaPt$_3$P.\\

\noindent\textbf{Discussion}\\
We investigate the normal state properties of LaPt$_3$P by a detailed band structure calculation using density functional theory within the generalized gradient approximation consistent with previous studies~\cite{Chen2012,Subedi2013}. LaPt$_3$P is centrosymmetric with a paramagnetic normal state respecting TRS. It has significant effects of spin-orbit coupling (SOC) induced band splitting near the Fermi level ($\sim 120$ meV, most apparent along the $MX$ high symmetry direction). Kramer's degeneracy survives in the presence of strong SOC due to centrosymmetry and SOC only produces small deformations in the Fermi surfaces~\cite{Yip2014}. The shapes of the Fermi surfaces play an important role in determining the thermodynamic properties of the material. The projections of the four Fermi surfaces of LaPt$_3$P on the $y-z$ and $x-y$ plane are shown in \fig{fig:SurfaceStates}\textbf{a} and \fig{fig:SurfaceStates}\textbf{b} respectively with the Fermi surface sheets having the most projected-DOS at the Fermi level shown in blue and orange. It shows the multi-band nature of LaPt$_3$P with orbital contributions mostly coming from the $5d$ orbitals of Pt and the $3p$ orbitals of P.

%Group theoretical classification of the SC order parameters within the Ginzburg-Landau theory~\cite{Sigrist1991,Ghosh2020} gives all possible SC instabilities based only on the symmetries of the normal state. 

LaPt$_3$P has a nonsymmorphic space group $P4/mmm$ (No. 129) with point group D$_{4h}$. From the group theoretical classification of the SC order parameters within the Ginzburg-Landau theory~\cite{Sigrist1991,Ghosh2020}, the only possible superconducting instabilities with strong SOC which can break TRS spontaneously at $T_c$ correspond to the two $2$D irreducible representations, $E_g$ and $E_u$, of D$_{4h}$~\cite{nonsymmnote1}. The superconducting ground state in the $E_g$ channel is a pseudospin \emph{chiral d-wave} singlet state with gap function $\Delta(\bk) = \Delta_0 k_z (k_x + i k_y)$ where $\Delta_0$ is an amplitude independent of $\bk$. While the $E_u$ order parameter is a pseudospin \emph{nonunitary chiral p-wave} triplet state with $d$-vector $\bd(\bk) = \left[ c_1 k_z, i c_1 k_z, c_2(k_x + i k_y)\right]$ where $c_1$ and $c_2$ are material dependent real constants independent of $\bk$.% nonzero in general. %These instabilities break additional crystalline symmetries necessarily requiring an unconventional pairing mechanism. 

% The normal state symmetry group of the system is given by $\mathcal{G} = G_0 \otimes U(1) \otimes \mathcal{T}$, where $U(1)$ is the gauge symmetry group, $G_0$ is the group of symmetries containing the point group symmetries of D$_{4h}$ and $SO(3)$ the spin rotation symmetries in $3$D, and $\mathcal{T}$ is the group of time-reversal symmetry (TRS). We construct the Ginzberg-Landau (GL) free energy of the system invariant under this symmetry group. The D$_{4h}$ point group has $8$ one dimensional irreducible representations (irreps) (4 of them has even parity and the other 4 has odd parity) and 2 two dimensional irreps (one with even parity E$_g$ and the other with odd parity E$_u$). Due to centrosymmetry this material has either purely triplet or purely singlet superconducting states in general. Furthermore, since a TRS breaking superconducting order parameter requires degenerate or multidimensional irreps, this system can have such type of instability only in the E$_g$ or the E$_u$ channels~\cite{Ghosh2020a}. 

We compute the quasi-particle excitation spectrum for the two TRS breaking states on a generic single band spherical Fermi surface using the Bogoliubov-de Gennes mean field theory~\cite{Sigrist1991,Ghosh2020}. The chiral $d$-wave singlet state leads to an energy gap $= |\Delta_0 ||k_z| \sqrt{k^2_x + k^2_y}$. It has a line node at the ``equator'' for $k_z = 0$ and two point nodes at the ``north'' and ``south'' poles (shown in \fig{fig:SurfaceStates}\textbf{a}). The low temperature thermodynamic properties are, however, dominated by the line node because of its larger low energy DOS than the point nodes. The triplet state has an energy gap $= \sqrt{g(k_x,k_y) + 2 c^2_1 k^2_z - 2 |c_1||k_z| \sqrt{f(k_x,k_y) + c^2_1 k^2_z}}$ where $f(k_x,k_y) = c^2_2 (k^2_x + k^2_y)$. It has only two point nodes at the two poles and no line nodes. Thus, the low temperature linear behaviour of the superfluid density of LaPt$_3$P shown in \fig{fig:sfdensity} is only possible in the chiral $d$-wave state with a line node in contrast to the triplet state with only point nodes which will give a quadratic behaviour and saturation at low temperatures. Thus LaPt$_3$P is one of the rare unconventional SCs for which we can unambiguously identify the superconducting order parameter. The point nodes and the line node for the chiral $d$-wave state on the Fermi surface sheets of LaPt$_3$P are shown in \fig{fig:SurfaceStates}\textbf{a} and \fig{fig:SurfaceStates}\textbf{b}.

%The inherent multiband nature of LaPt$_3$P with significant effect of SOC near the Fermi level together with the spontaneously broken TRS at $T_c$ and centrosymmetry implies, in general, that for the chiral d-wave ground state the two point nodes at the two poles and the line node at the equator get inflated to form Bogoliubov Fermi surfaces (BFSs)~\cite{Agterberg2017,Brydon2018} (\ttds{as shown in fig?}). The inflation of the nodes is caused by a pseudo-magnetic field of strength $h$ which is of second order in interband coupling. Thus it is in general small but non zero quantity and we consider $\mu_0 h \ll k_B T_c$ with $\mu_0$ being the vacuum permeability. These BFSs are topologically protected by a $Z_2$ topological invariant and in turn energetically stabilize the superconducting ground state with SOC providing additional stability~\cite{Agterberg2017,Menke2019}. The low-energy quasi-particle density of states (qpDOS) gets affected qualitatively by the presence of the BFSs~\cite{Lapp2020}. As a result, the low temperature behaviour of the superfluid density ($\rho_s$) measured by the TF-field $\mu$SR for an inflated line node changes to $\left[\frac{\rho_s(T)}{\rho_s(0)} - 1 \right] \propto - (k_B T) e^{-\frac{\mu_0 h}{k_B T}}$ but an inflated point node has the same effect as a normal point node. Our data shown in \fig{fig:lambdaT} does not exclude the presence of inflated line nodes but since $h$ is very small it is difficult to clearly distinguish it from an ordinary line node.

We now discuss the topological properties of the chiral $d$-wave state of LaPt$_3$P based on a generic single-band spherical Fermi surface (chemical potential $\mu = k^2_F/(2m)$ where $k_F$ is the Fermi wave vector and $m$ is the electron mass)~\cite{Goswami2013,Schnyder2015}. However, topological protection of the nodes also ensures stability against multiband effects. The effective angular momentum of the Cooper pairs is $L_z = +1$ (in units of $\hbar$) with respect to the chiral $c$-axis. The equatorial line node acts as a vortex loop in momentum space~\cite{heikkila2011} and is topologically protected by a $1$D winding number $w(k_x,k_y) = 1$ for $k^2_x+k^2_y <k^2_F$ and $=0$ otherwise. The nontrivial topology of the line node leads to two-fold degenerate zero-energy Majorana bound states in a flat band on the $(0,0,1)$ surface BZ as shown in \fig{fig:SurfaceStates}\textbf{c}. As a result, there is a diverging zero-energy DOS leading to a zero-bias conductance peak (which can be really sharp~\cite{Kobayashi2015}) measurable in STM. This inversion symmetry protected line node is extra stable due to even parity SC~\cite{Kobayashi2014,Kobayashi2015}. The point nodes on the other hand are Weyl nodes and are impossible to gap out by symmetry-preserving perturbations. They act as a monopole and an anti-monopole of Berry flux as shown in \fig{fig:SurfaceStates}\textbf{d} and are characterized by a $k_z$ dependent topological invariant, the sliced Chern number $C(k_z) = L_z$ for $|k_z|<k_F$ with $k_z \neq 0$ and $=0$ otherwise (see the Supplemental Material for details). As a result, the $(1,0,0)$ and $(0,1,0)$ surface BZs each have a Majorana Fermi arc which can be probed by STM as shown in \fig{fig:SurfaceStates}\textbf{c}. There are two-fold degenerate chiral surface states with linear dispersion carrying surface currents leading to local magnetisation that may be detectable using SQUID magnetometry. One of the key signatures of chiral edge states is the anomalous thermal Hall effect (ATHE) which depends on the length of the Fermi arc in this case. Impurities in the bulk can, however, increase the ATHE signal by orders of magnitude~\cite{Ngampruetikorn2020} over the edge contribution making it possible to detect with current experimental technology \cite{Hirschberger2015}. We also note that a $90^\circ$ rotation around the $c$-axis for the chiral $d$-wave state leads to a phase shift of $\pi/2$ which can be measured by corner Josephson junctions~\cite{Strand2009}.

%To further establish the nature of the chiral $d$-wave state and the presence of the Bogoliubov Fermi surfaces it is important to carefully measure upto a very low temperature the observables such as electronic specific heat and the NMR relaxation~\cite{Lapp2020}. Impurity scattering plays a central role in determining order parameter symmetries in nodal USCs. In this regard, a rapidly developing technique is the phase sensitive quasi-particle interference imaging using scanning tunnelling microscopy (STM)~\cite{Allan2013,Jakob2020}. We expect definitive experimental signatures to be realized in this technique corresponding to the sign-changing chiral d-wave state realized in LaPt$_3$P.  

%\section{Conclusions}
%In summary we have used $\mu$SR technique, first principles band structure computation, symmetry analysis based on Ginzburg-Landau theory and Bogoliubov de-Gennes mean field theory to characterize the unconventional superconducting properties of LaPt$_3$P. The phenomenon of TRS breaking pairing state measured by ZF-$\mu$SR and the signature of line nodes in the superfluid density measured by the TF-$\mu$SR can be explained well by the symmetry allowed spin-singlet chiral $d$-wave order parameter. Topologically protected BFSs are realized in this material due to multiband nature and strong SOC effect. The unconventional properties of TRS breaking and nodal gap structure together with its topologically protected BFSs make LaPt$_3$P a unique platform for the realization and exploration of Majorana fermions.     

\textit{Acknowledgments}: PKB gratefully acknowledges the ISIS Pulsed Neutron and Muon Source of the UK Science \& Technology Facilities Council (STFC) and Paul Scherrer Institut (PSI) in Switzerland for access to the muon beam times.  SKG thanks Jorge Quintanilla and Adhip Agarwala for stimulating discussions and acknowledges the Leverhulme Trust for support through the Leverhulme early career fellowship. The work at the University of Warwick was funded by EPSRC,UK, Grant EP/T005963/1. XX is partially supported by the National Natural Science Foundation of China (Grants 11974061, U1732162). NDZ thanks K. Povarov and acknowledges support from the Laboratory for Solid State Physics, ETH Zurich where synthesis studies were initiated.

\section{Supplementary material}

In this Supplemental Material, we present details of the synthesis, characterization measurements, experimental methods and data analysis of the LaPt$_{3}$P samples grown at Warwick, United Kingdom and at ETH, Switzerland. We also give additional band structure results, details of the symmetry analysis and topological properties of the chiral $d$-wave state of LaPt$_3$P.

\subsection{Synthesis and characterization of the sample grown at Warwick, United Kingdom}

Polycrystalline LaPt$_{3}$P~\cite{Takayama2012} samples (sample-A) were synthesized by a solid state reaction method. Powders of elemental platinum, red phosphorus, and alkaline earth (lanthanum) were mixed in an argon-filled glove
box, and sealed in a quartz tube filled with argon gas. The tube was initially heated to $400$ $^{\circ}$C and held at this temperature for 12 h in order to avoid rapid volatilization of phosphorus, then reacted at $900$ $^{\circ}$C for 72 h. The sintered pellet was reground and further annealed at $900$ $^{\circ}$C within argon-filled quartz tubes for several days and finally quenched into iced water.

\begin{figure}[htb]
\begin{center}
\includegraphics[width=0.99\columnwidth]{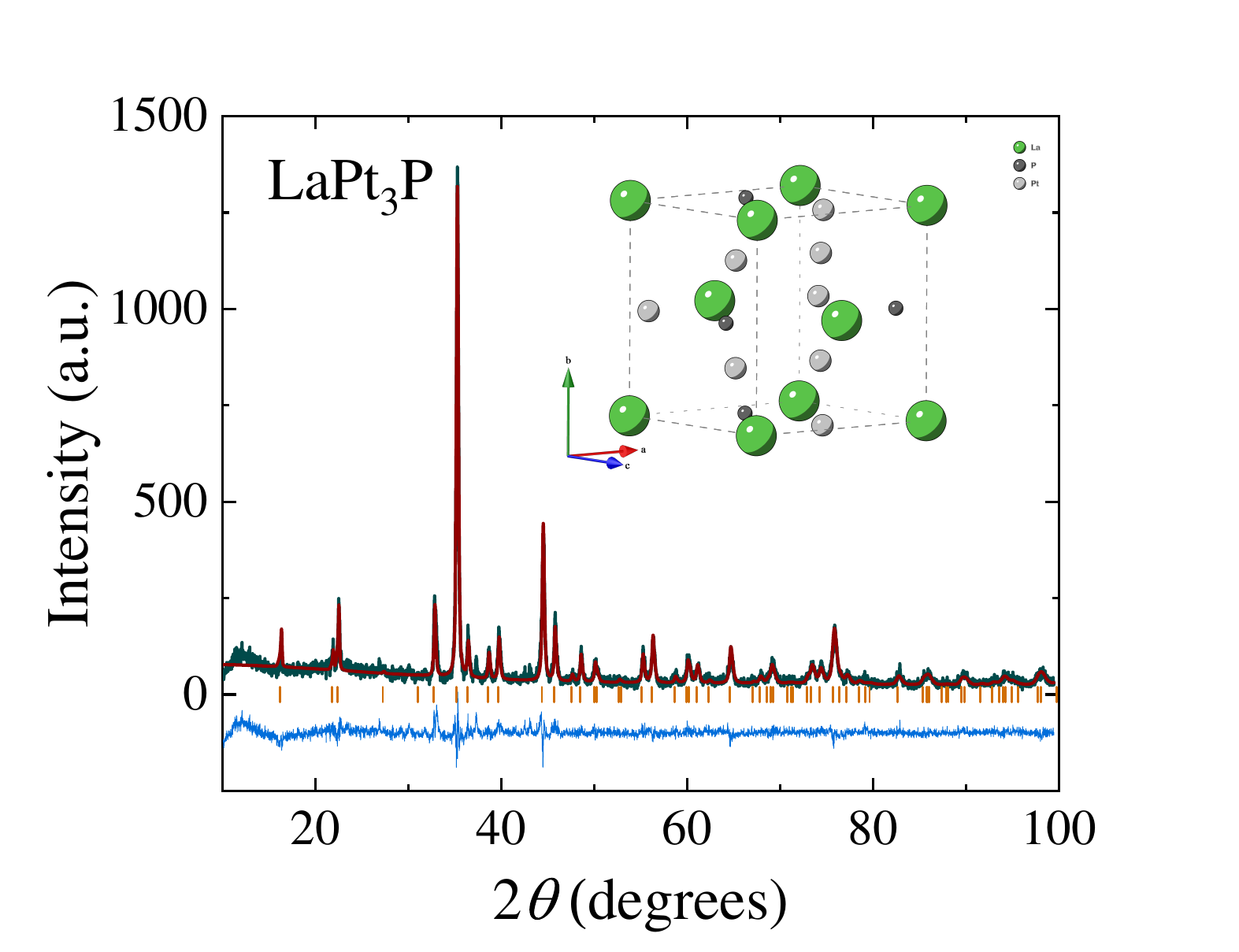}
\vspace{-0.25cm}
\caption{\textbf{Powder x-ray diffraction pattern of the sample-A of LaPt$_3$P at room temperature.} X-ray diffraction pattern of LaPt$_3$P at room temperature where the green, red and blue lines indicate the experimental data, the fit and the difference between the data and the fit, respectively. The orange dashes indicate the expected Bragg peaks. The inset shows the structure of a unit cell of LaPt$_3$P.}
\label{fig:PXRD_UK}
\end{center}
\end{figure}

The room-temperature structure was determined via powder x-ray diffraction (PXRD). PXRD was measured using a Bruker D5000 general purposed powder diffractometer. The diffraction pattern is shown in \fig{fig:PXRD_UK}. Reitveld refinement was carried out using the TOPAS software package \cite{Coelho2018} which gave the parameters shown in Table I.%\tab{tab:structure}.

\begin{table}[h]
\mylabel{tab:structure}
\caption{Crystallographic and Rietveld refinement parameters obtained on LaPt$_{3}$P.}
\begin{center}
\begin{tabular}{|l l|} 
 \hline 
  Space-group & $P4nmm$ (No. 129) \\
  Formula units/unit cell (Z) & 2 \\
  Lattice parameter & \\
  $a (\mathrm{\AA})$ & $5.7683(6)$ (at 300K) \\
  $c (\mathrm{\AA})$ & $5.4681(7)$ (at 300K) \\
  $V_{\mathrm{cell}} (\mathrm{\AA}^3)$ & $182.4$ \\
 \hline
\end{tabular}
\hspace{40pt}
\begin{tabular}{|l l l l l l|} 
 \hline 
Atom & Wyckoff Position & Occupancy & x & y & z  \\
 \hline
Pt(1) & 4e & 1 & 0.25 & 0.25 & 0.5  \\
Pt(2) & 2c & 1 & 0 & 0.5 & 0.1476(8) \\
La(1) & 2a & 1 & 0 & 0.5 & 0.758(4) \\
P(1) & 2c & 1 & 0 & 0 & 0 \\
 \hline \hline
\end{tabular}
\end{center}
\end{table}

\begin{figure}[htb]
\begin{center}
\includegraphics[width=0.85\columnwidth]{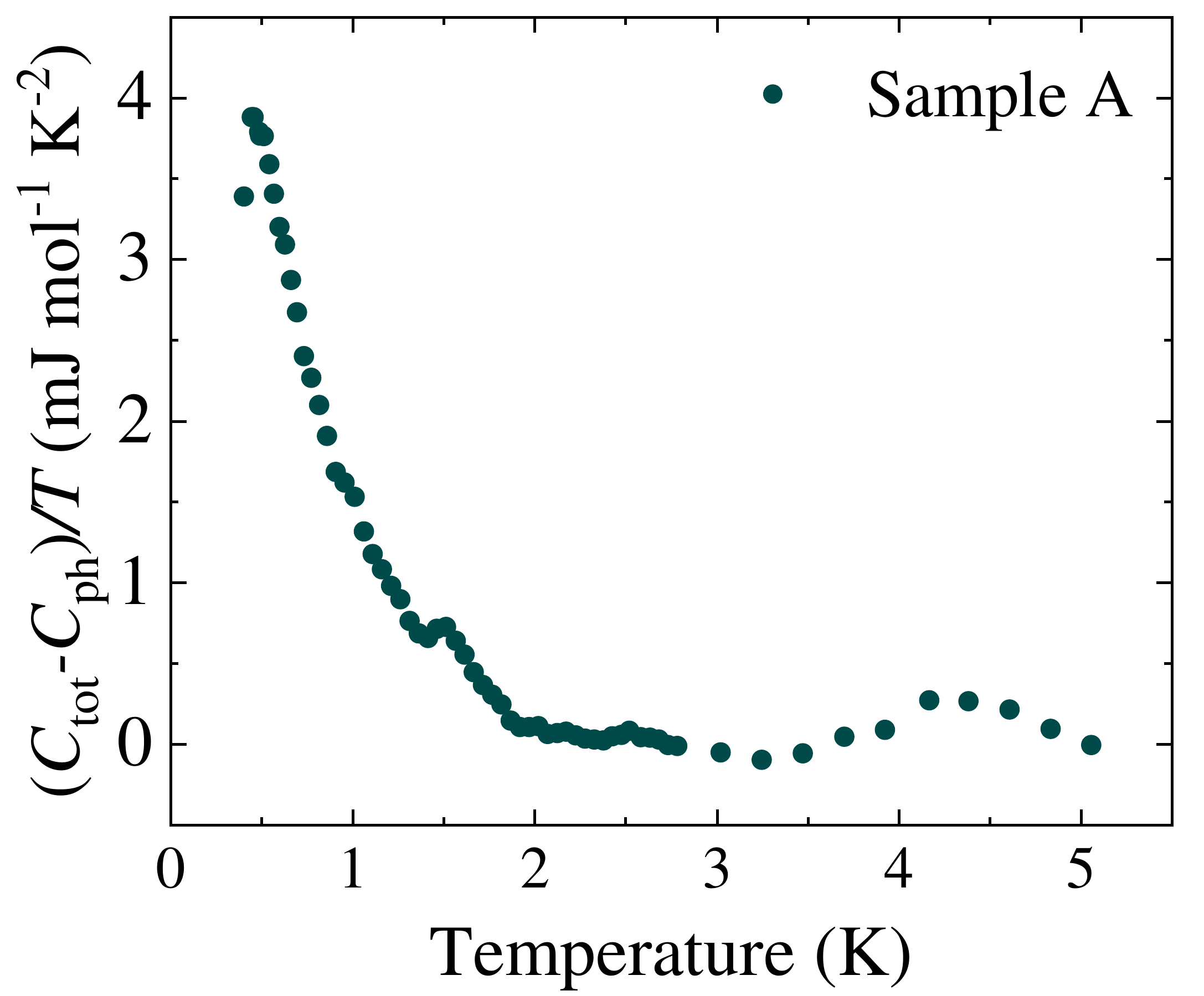}
\vspace{-0.25cm}
\caption{\textbf{Heat capacity of the sample-A of LaPt$_3$P in zero-field.} $(C_{tot} - C_{ph})/T$ as a function of temperature. There is a small anomaly close to the expected superconducting transition temperature that is masked by a large hyperfine contribution.}
\label{fig:spheat_UK}
\end{center}
\end{figure}

The heat capacity in zero field was measured using a Quantum Design Physical Property Measurement System (PPMS) with a He$^3$ insert to get down to 0.5 K. The total specific heat $C_{tot}$  at low temperatures is made up of several contributions,
\beq
C_{tot} =  C_{el} + C_{ph} + C_{hyp}
\eeq
where $C_{el}$ is the electronic specific heat having the form in the normal state
\beq
C_{el} = \gamma_n T
\eeq
with $\gamma_n$ being the Sommerfeld coefficient, $C_{ph}$ is the specific heat due to the phonons given by 
\beq
C_{ph} = \beta_3 T^3 + \beta_5 T^5
\eeq
with $\beta_3$ and $\beta_5$ being temperature independent parameters, and $C_{hyp}$ is a contribution due to hyerfine splitting
\beq
C_{hyp} \propto 1/T^2.
\eeq

Fitting the normal state specific heat gives $\gamma_n = 9.78(7)$ mJ/mol-K, $\beta_3 = 0.369(14)$ mJ/mol-K$^4$ and $\beta_5 = 5.47(5)$ $\mu$J/mol-K$^4$. We then subtract the phonon contribution to the specific heat to plot the electronic specific heat including the hyperfine contribution. This is shown in the \fig{fig:spheat_UK} for the sample-A of LaPt$_3$P, and is consistent with the previous measurement of Ref.\cite{Takayama2012}. We note that the specific heat has a small anomaly close to the expected $T_c \approx 1.1$ K which is obscured by an upturn at lower temperatures. This is due to a large hyperfine contribution to the specific heat.

\begin{figure}[htb]
\begin{center}
\includegraphics[width=0.95\columnwidth]{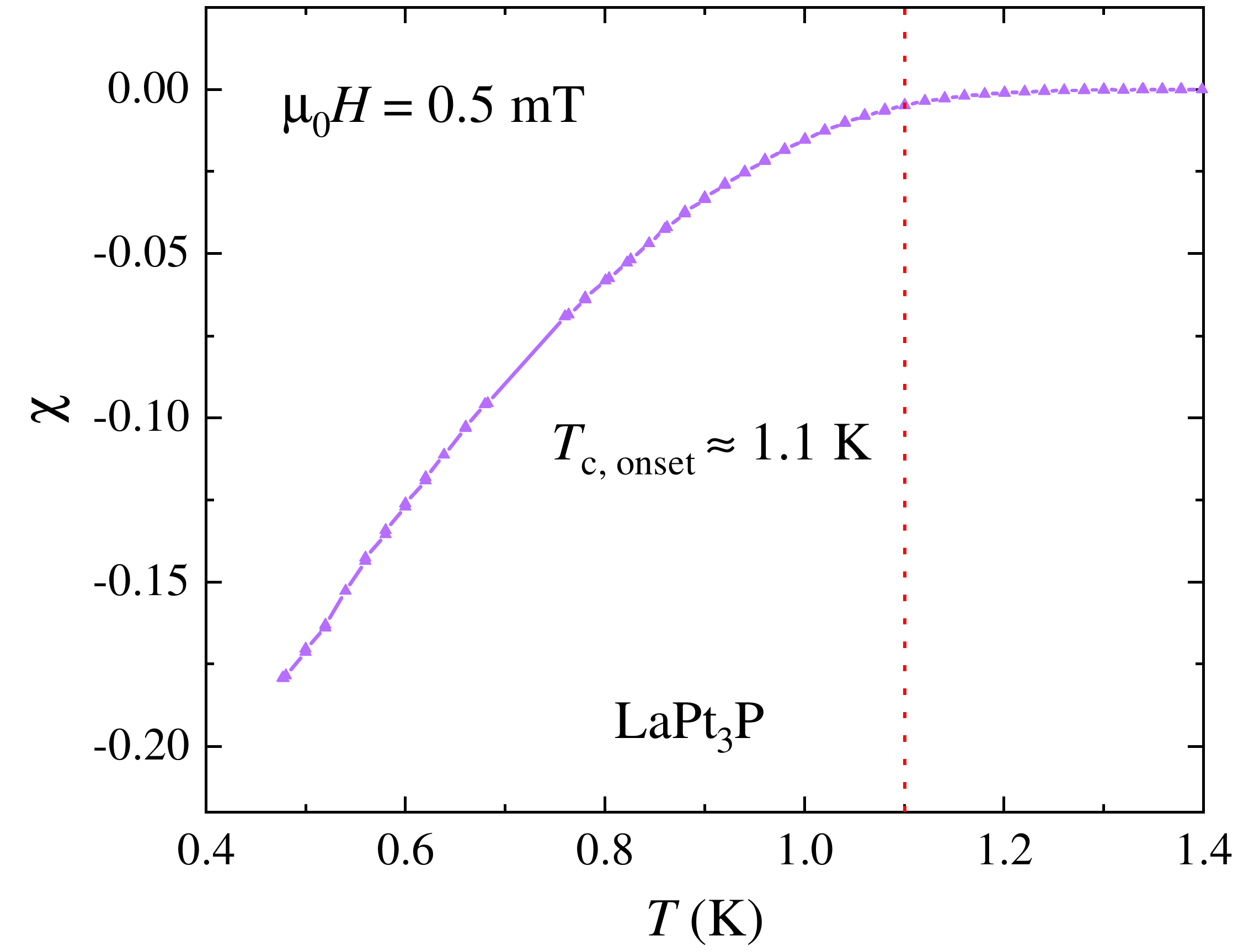}
\vspace{-0.25cm}
\caption{\textbf{Zero-field-cooled magnetic susceptibility of the sample-A of LaPt$_3$P as a function of temperature.}}
\label{fig:susceptibility_UK}
\end{center}
\end{figure}

The magnetic susceptibility was measured using a Quantum Design Magnetic Property Measurement System (MPMS) using an i-quantum $^3$He insert. As seen from \fig{fig:susceptibility_UK} this sample has a relatively low Meissner fraction ($\sim 30 \%$).

\subsection{Synthesis and characterization of the sample grown at ETH, Switzerland}

A polycrystalline sample of LaPt$_{3}$P (sample-B) was synthesized using the cubic anvil high-pressure and high-temperature technique. Starting powders of LaP and Pt of high purity (99.99\%) were weighed according to the stoichiometric ratio, thoroughly ground, and enclosed in a boron nitride container, which was placed inside a pyrophyllite cube with a graphite heater. The details of experimental setup can be found in Ref.\cite{ZHIGADLO201694}. All the work related to the sample preparation and the packing of the high pressure cell-assembly was performed in an argon-filled glove box. In a typical run, a pressure of 2 GPa was applied at room temperature. The temperature was ramped in 3 h to the maximum value of 1500\,$^\circ$C, maintained for 5 h, and then cooled to 1350\,$^\circ$C over 5 h and finally reduced to room temperature in 3 h. Afterward, the pressure was released, and the sample was removed. The sample exhibits a large diamagnetic response with the superconducting transition temperature of 1.1 K.

Susceptibility measurements were performed using a Quantum Design Magnetic Property Measurement System (MPMS) by cooling the sample at base temperature in zero field and then apply 7 mT magnetic field. Data were collected while warming up the sample temperature. As shown in the main text, the temperature dependence of the susceptibility data shows a bulk superconducting transition with a $T_{\rm c}$ at around 1.1 K.

%\begin{figure}[htb]
%\begin{center}
%\includegraphics[width=0.55\columnwidth]{susceptibility_ETH.pdf}
%\vspace{-0.25cm}
%\caption{(Color online) Temperature dependence of the susceptibility data showing bulk superconducting transition at 1.1K of the sample grown at ETH.}
% \label{fig:susceptibility_ETH}
%\end{center}
%\end{figure}

\begin{figure}[htb]
\begin{center}
\includegraphics[width=0.95\columnwidth]{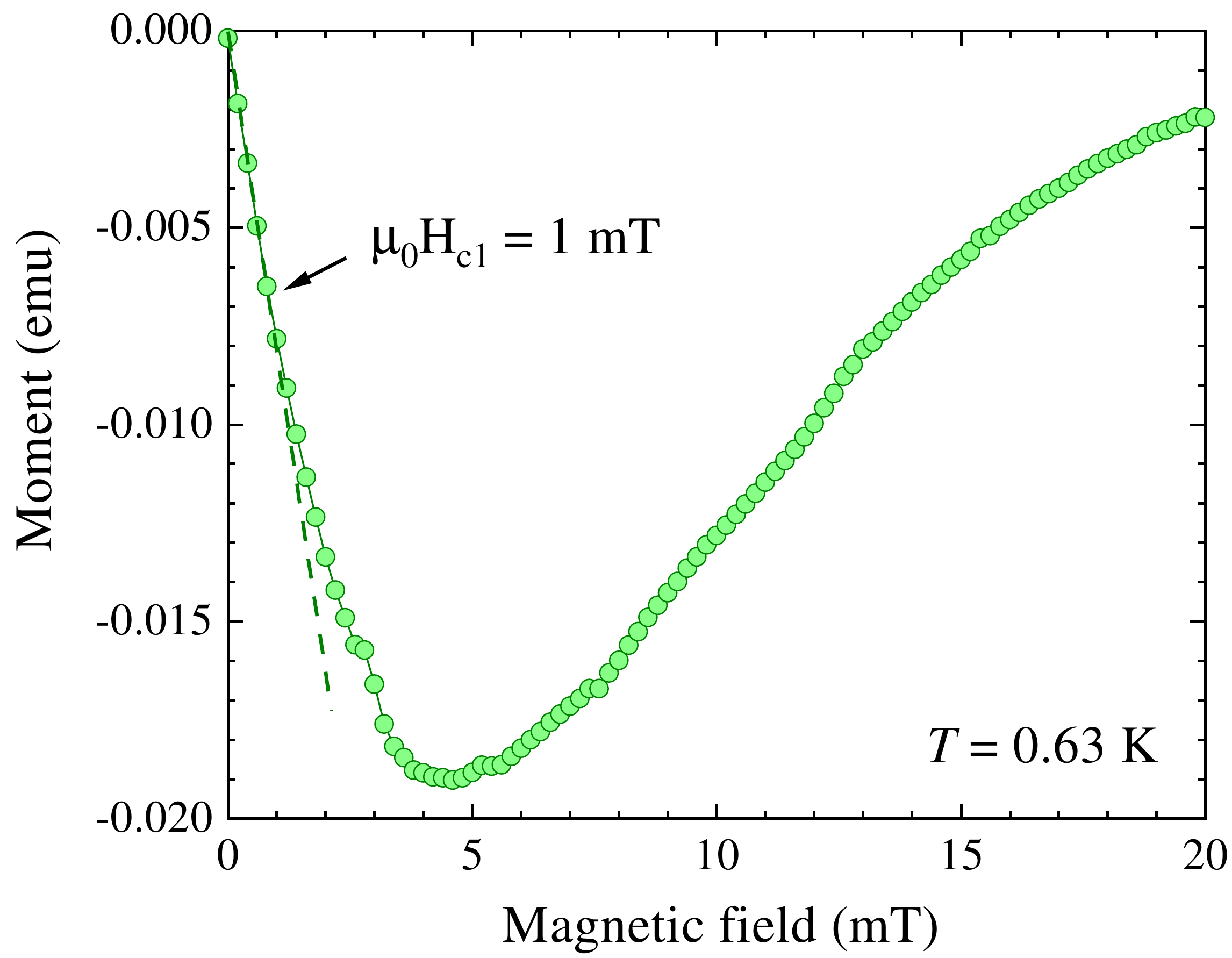}
\vspace{-0.25cm}
\caption{\textbf{Magnetic field dependence of the virgin magnetisation curve for sample-B of LaPt$_{3}$P.} We note that the lower critical field $\mu_0 H_{\rm c1} \approx 1$ mT.}
\label{fig:MvH_ETH}
\end{center}
\end{figure}

A virgin magnetisation curve was measured at 0.63 K in a Quantum Design MPMS. A linear deviation of the magnetisation curve at low field region (see \fig{fig:MvH_ETH}) shows that the lower critical field $H_{\rm c1}$ of LaPt$_{3}$P is around 1 mT.

\subsection{$\mu$SR technique}

$\mu$SR is a very sensitive local magnetic probe utilizing fully spin-polarized muons~\cite{yaouanc2011}. In a $\mu$SR experiment polarized muons are implanted into the host sample. After thermalization, each implanted muon decays (lifetime $\tau_\mu = 2.2$ $\mu$s) into a positron (and two neutrinos) emitted preferentially in the direction of the muon's spin at the time of decay. Using detectors appropriately positioned around the sample, the decay positrons are detected and time stamped. From the collected histograms, the asymmetry in the positron emission as a function of time, $A(t)$, can be determined, which is
directly proportional to the time evolution of the muon spin polarization. 

$\mu$SR measurements were performed on sample-A in the MUSR spectrometer at the ISIS Pulsed Neutron and Muon Source, UK, and  on sample-B in the LTF spectrometer at the Paul Scherrer Institut (PSI), Switzerland. The polycrystalline samples of LaPt$_{3}$P in the form of powder were mounted on high purity silver sample holders. The samples were cooled from above $T_{\rm c}$ to base temperature in zero field for ZF-$\mu$SR measurements, and in a field for the TF-$\mu$SR measurements. The external field was 10~mT for the TF-$\mu$SR measurements performed at ISIS and was 7~mT for the TF-$\mu$SR measurements performed at PSI. ZF-$\mu$SR measurements were performed in true zero field, achieved by three sets of orthogonal coils working as an active compensation system which cancel any stray fields at the sample position down to 1.0 $\mu$T. LF-$\mu$SR measurements were also performed under similar field-cooled conditions. The typical counting statistics were $\sim40$ and $\sim24$ million muon decays per data point at ISIS and PSI, respectively. The ZF-, LF- and TF-$\mu$SR data were analyzed using the equations given in the text. 

The zero temperature upper critical field for LaPt$_{3}$P, $\mu_0 H_{c2} \approx 0.12$ T which is much larger than the applied transverse fields in the TF-$\mu$SR measurements. The detailed parameters for the analysis of superfluid density data from the TF-$\mu$SR measurements for the two samples using the different gap models mentioned in the main text are given in the Table II.%\tab{tab:sfden}.

\begin{table*}[htb]
\mylabel{tab:sfden}
\caption{Summary of the analysis of the superfluid density data for the two samples of LaPt$_{3}$P.}
\begin{center}
\begin{tabular}{l l l l l l l l l l l l l} 
 \hline 
Model &&g($\theta, \phi$) && Gap type && Reduced least-squared deviation ($\chi^2_r$) && Fitted $\Delta_m(0)/(k_B T_c)$   \\
\hline
$s$-wave && 1 && nodeless && \hspace{2cm} $13.025$ && \hspace{0.5cm} $1.270 \pm 0.020$   \\
$p$-wave && $\sin(\theta) e^{i\phi}$ && two point nodes && \hspace{2cm} $4.537$  && \hspace{0.5cm} $1.693 \pm 0.029$ \\
chiral $d$-wave && $\sin(2\theta) e^{i\phi}$ && two point nodes + a line node && \hspace{2cm} $2.238$ && \hspace{0.5cm} $1.989 \pm 0.011$ \\
\hline \hline
\end{tabular}
\end{center}
\end{table*}

\subsection{Band structure}
LaPt$_3$P crystallizes in a centrosymmetric primitive tetragonal crystal structure. The corresponding space group is P4/nmm (No. 129) which is nonsymmorphic. The point group of the Bravais lattice is D$_{4h}$. The nonsymmorphic symmetries within a unit cell include both screw axes and glide planes. We have performed detailed band structure calculations of LaPt$_3$P using density functional theory (DFT). The corresponding band structure results with and without spin orbit coupling (SOC) are shown in \fig{fig:bands}(a) and \fig{fig:bands}(b) respectively. We note that this material has significant splitting of bands due to SOC~\cite{Chen2012}. The maximum band splitting caused by the SOC near the Fermi level is estimated to be $\sim 120$ meV and is most apparent along the MX high symmetry direction. The SOC induced band splitting breaks the spin-symmetry and have important consequences in Cooper-pairing in this material. 

The 3D Fermi surfaces were ploted by the \texttt{XCrySDen} packages~\cite{Kokalj2003}. The Fermi surfaces with SOC are shown in \fig{fig:Fermi_surface}. We note that there are four Fermi surfaces with the middle two shown in \fig{fig:Fermi_surface}(b) and \fig{fig:Fermi_surface}(c); and again in \fig{fig:Fermi_surface}(f) and \fig{fig:Fermi_surface}(g) from a different view, contributing the most to the density of states (DOS) at the Fermi level. This is seen from the projected DOS at the Fermi level shown in \fig{fig:pdos}. \fig{fig:pdos}(a) shows the contributions of the different atomic orbitals to the DOS at the Fermi level. We note that Pt-5d orbitals contribute the most. Thus LaPt$_3$P is a multi-band system. \fig{fig:pdos}(a) shows the contributions of the different Fermi surfaces to the DOS at the Fermi level.

\begin{figure*}[!hbt]
\includegraphics[width=0.90\textwidth]{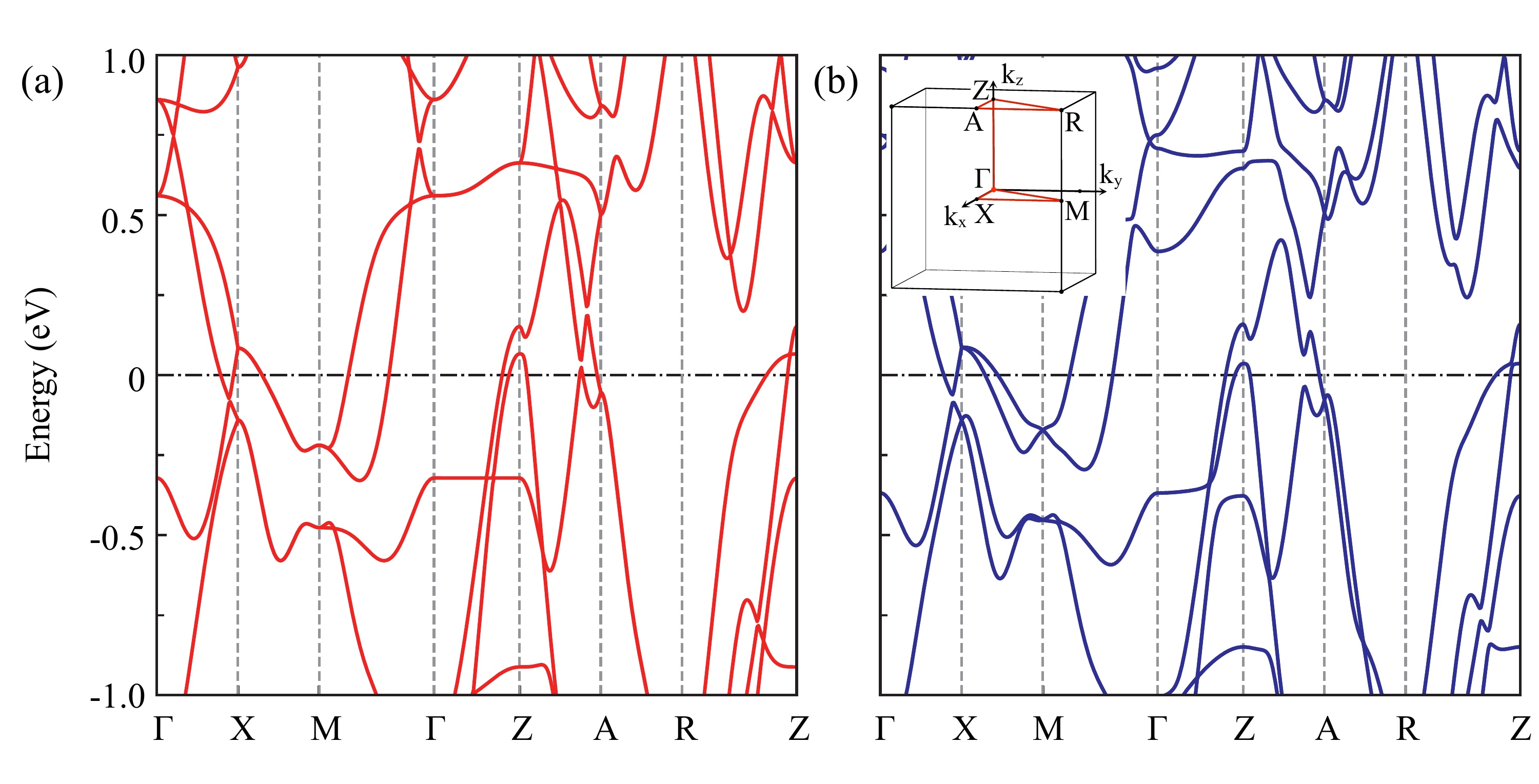}
\caption{\textbf{First principles band structure results of LaPt$_3$P.} a) Band structure without SOC. b) Band structure with SOC. The primitive tetragonal Brillouin zone with the marked high symmetry points and directions used in the band structure computation is shown in the inset of (b). We note that SOC induces significant band splitting near the Fermi level especially from M to X.}
\mylabel{fig:bands}
\end{figure*}

\begin{figure*}[!hbt]
\includegraphics[width=0.98\textwidth]{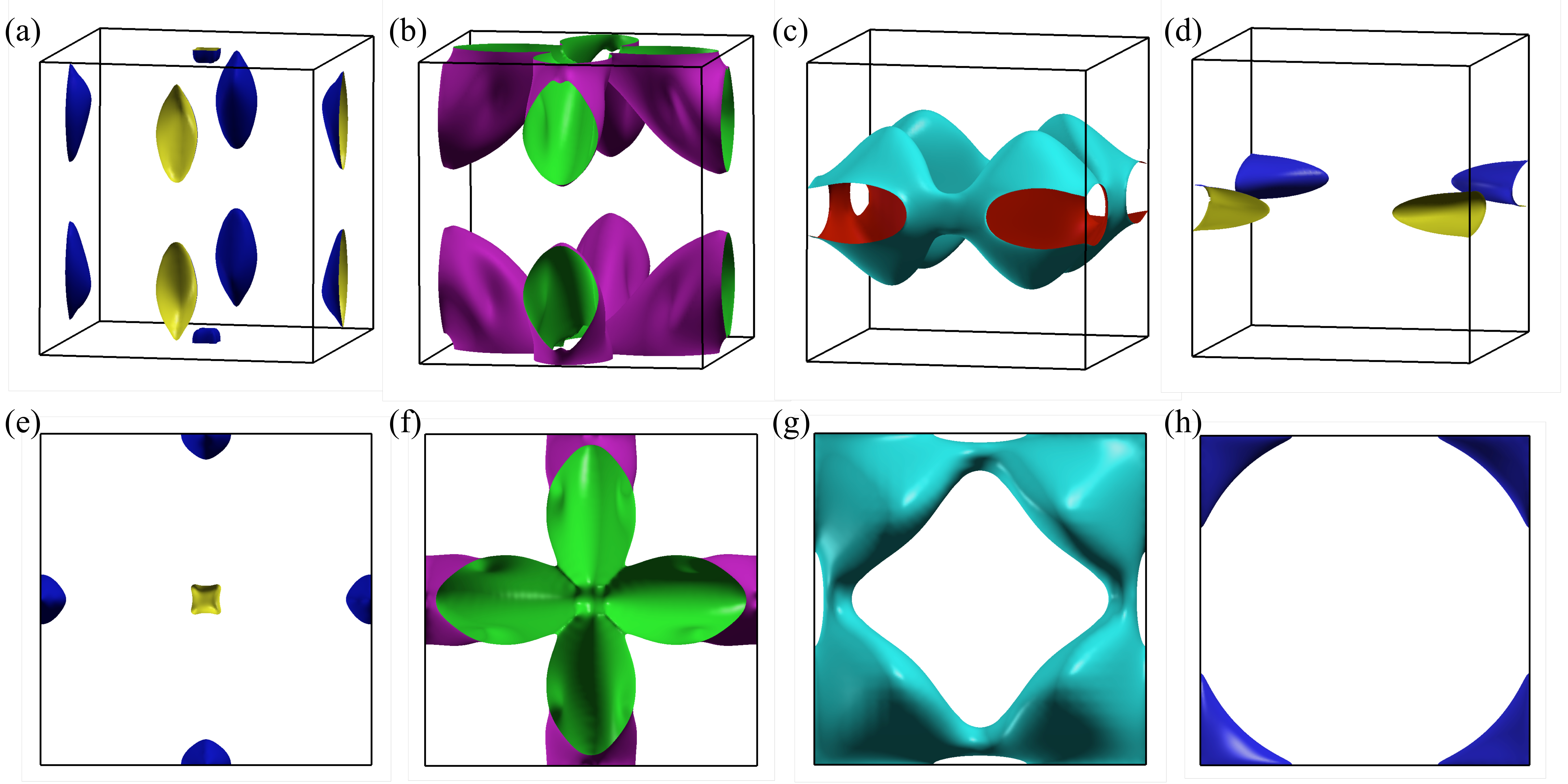}\\
\includegraphics[width=0.30\textwidth]{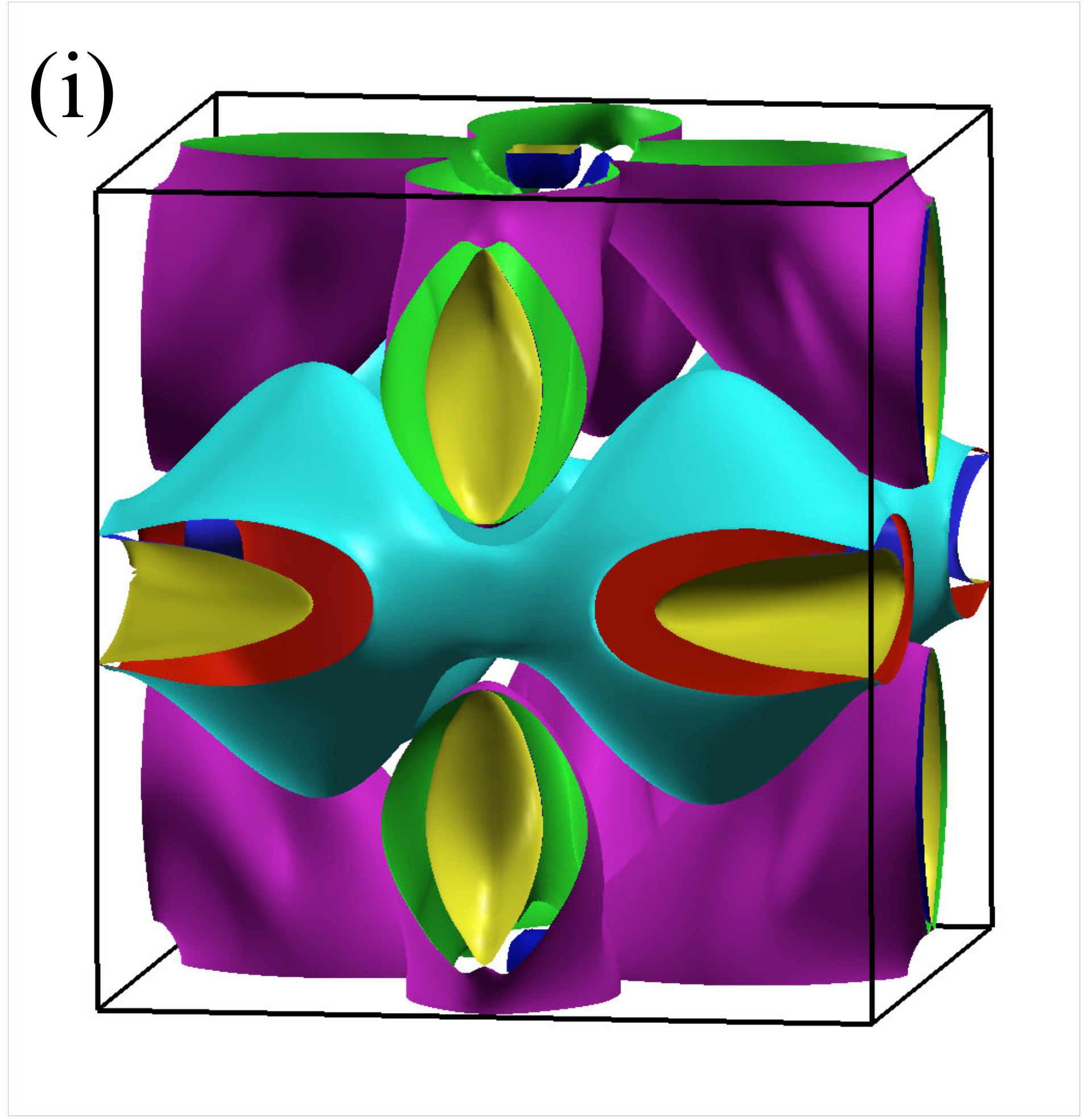}
\caption{\textbf{Fermi surfaces of LaPt$_3$P with SOC.} Panels (a)--(d) are from a side view and the panels (e)--(h) are from the top view for the four Fermi surface sheets and (i) shows a combined Fermi surface.}
\mylabel{fig:Fermi_surface}
\end{figure*}

\begin{figure*}[!hbt]
\includegraphics[width=0.90\textwidth]{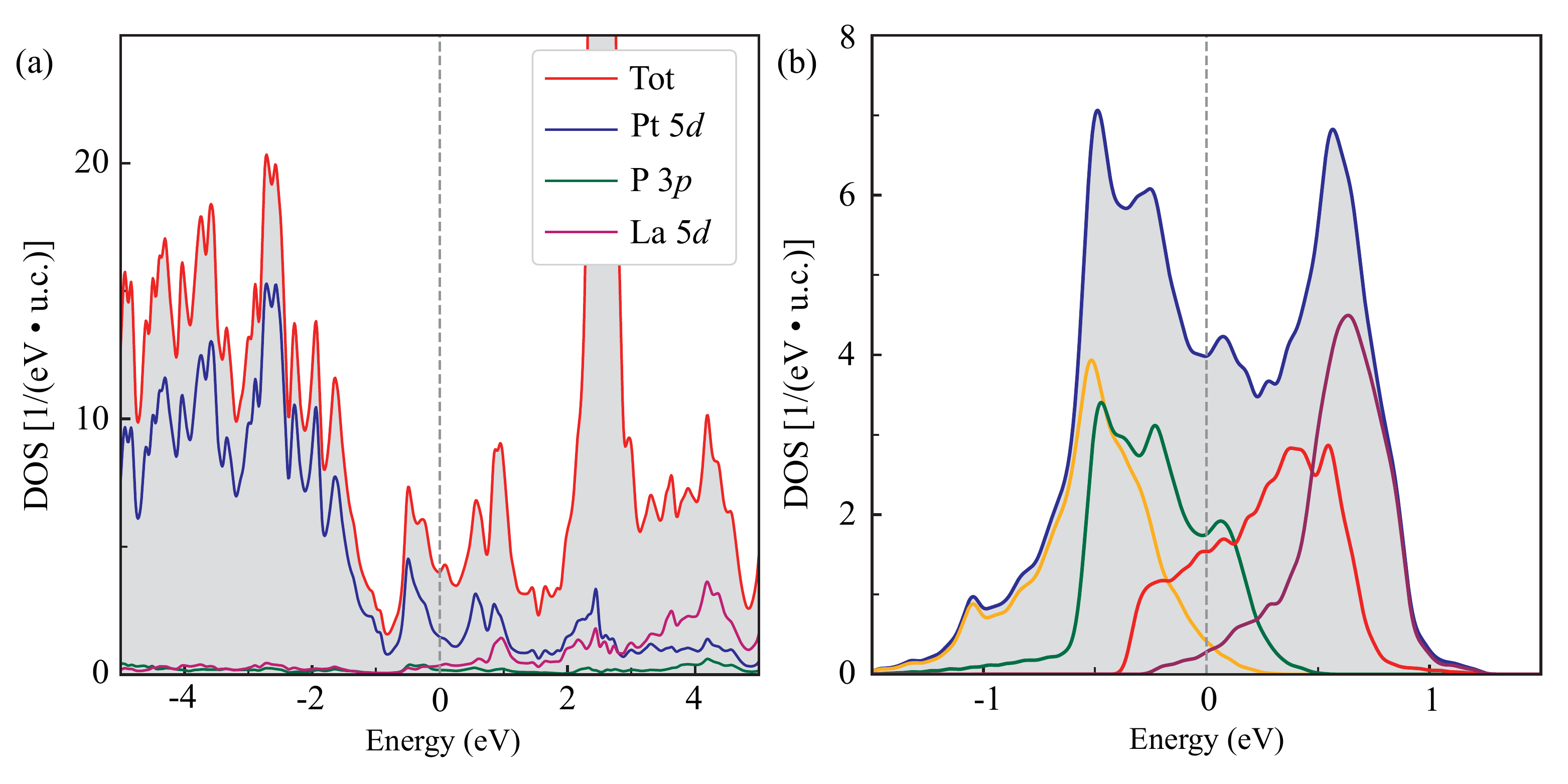}
\caption{\textbf{Projected density of states (DOS) results.} Left panel shows the contributions of different orbitals to the DOS. We note that Pt 5d orbitals contribute the most to the DOS at the Fermi level. The right panel shows the DOS contributions of the different Fermi surfaces. The blue is total and the other four correspond to the four Fermi surfaces. Their contributions at the Fermi level are 10.3\%, 43.4\%, 39.5\% and 6.3\%.}
\mylabel{fig:pdos}
\end{figure*}

\subsection{Symmetry analysis}
In this section we describe, the symmetry analysis of the possible superconducting order parameters for LaPt$_3$P. To proceed, we note the properties of the material: it is centrosymmetric, has nonsymmorphic symmetries, has considerable effects of SOC, has multiple bands potentially participating in superconductivity, has spontaneously broken TRS at $T_c$ and has line nodes dominating its thermodynamic behavior.

The normal state symmetry group of the system is given by $\mathcal{G} = G_0 \otimes U(1) \otimes \mathcal{T}$, where $U(1)$ is the gauge symmetry group, $G_0$ is the group of symmetries containing the point group symmetries of D$_{4h}$ and spin rotation symmetries in $3$D of $SO(3)$ and $\mathcal{T}$ is the group of time-reversal symmetry (TRS). The Ginzburg-Landau (GL) free energy of the system must be invariant under this symmetry group.

The D$_{4h}$ point group has $8$ one-dimensional irreducible representations (irreps) (4 of them have even parity and the other 4 have odd parity) and 2 two dimensional irreps (one with even parity denoted by $E_g$ and the other with odd parity denoted by $E_u$). Centrosymmetry implies that this material has either purely triplet or purely singlet superconducting instability in general. Furthermore, a TRS breaking superconducting order parameter requires degenerate or multi-dimensional irreps. This system can thus lead to such type of instability only in the $E_g$ or the $E_u$ irrep. We will now focus only on these two irreps and construct possible superconducting order parameters for the system. We consider strong SOC as uncovered by the band structure calculation of this material. 

The fourth order invariant corresponding to the 2 two-dimensional irreps $E_g$ and $E_u$ of $D_{4h}$ gives the quartic order term of the GL free energy~\cite{Annett1990,Sigrist1991} to be
\beq\mylabel{eq:fenergy}
f_4 = \beta_1 (|\eta_1|^2 + |\eta_2|^2)^2 + \beta_2 |\eta^2_1 + \eta^2_2|^2 + \beta_3 (|\eta_1|^4 + |\eta_2|^4)
\eeq
where $(\eta_1, \eta_2)$ are the two complex components of the two-dimensional order parameters. This free energy needs to be minimized with respect to both $\eta_1$ and $\eta_2$. The nonequivalent solutions are: $(\eta_1,\eta_2) = (1,0)$, $\frac{1}{\sqrt{2}}(1,1)$ and $\frac{1}{\sqrt{2}}(1,i)$. There is an extended region in the parameter space where the states corresponding to $(\eta_1,\eta_2) = (1,i)$ is stabilized. The instabilities corresponding  to this case spontaneously break TRS at $T_c$ due to a nontrivial phase difference between the two order parameter components.

Then the even parity superconducting order parameter belonging to $E_g$ is given by
\beq\mylabel{eqn:singlet}
\Delta(\bk) = \Delta_0 k_z (k_x + i k_y)
\eeq
where $\Delta_0$ is the real amplitude independent of $\bk$. This is a \emph{chiral d-wave} singlet order parameter. The odd parity superconducting order parameter belonging to $E_u$ gives rise to the gap matrix $\hat{\Delta}(\bk) = [\bd(\bk).\vec{\sigma}]i\sigma_y$ where $\vec{\sigma}$ denotes the three Pauli spin matrices and $\bd(\bk)$ is the triplet $d$-vector given by 
\beq\mylabel{eqn:triplet}
\bd(\bk) = \left[ A k_z, i A k_z, B(k_x + i k_y)\right].
\eeq 
Here, $A$ and $B$ are material dependent real constants independent of $\bk$ and in general they are nonzero. We note that the values of $A$ and $B$ determine the orientation of the $d$-vector. For example, for $A=0$ the $d$-vector points along the $c$-axis and for $B=0$ the $d$-vector points in the $ab$-plane. We also note that
\beq
\bd(\bk) \times \bd^*(\bk) = 2 i A k_z (B k_x \hat{x} - B k_y \hat{y} - A k_z \hat{z})
\eeq 
which is nonzero in general. Hence, this superconducting state is \emph{nonunitary chiral p-wave} triplet state.

\begin{figure*}[htb]
{
\centerline{
\includegraphics[width=0.25\textwidth]{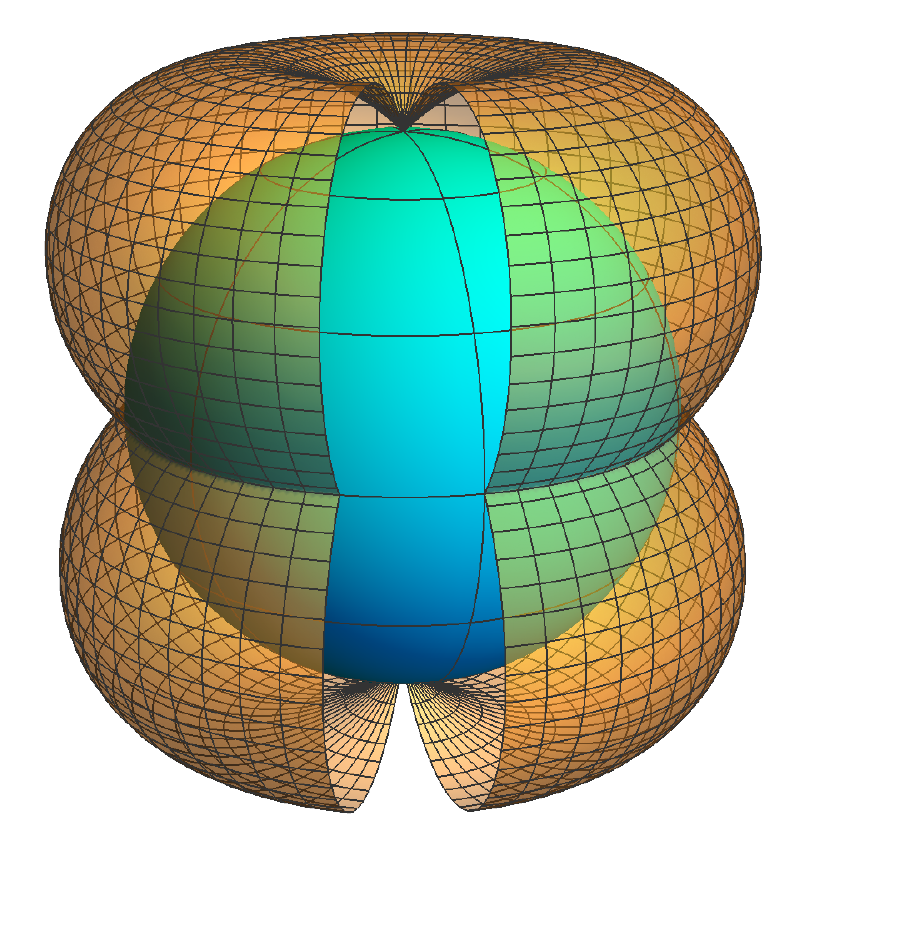}~~~~~~~~\includegraphics[width=0.35\textwidth]{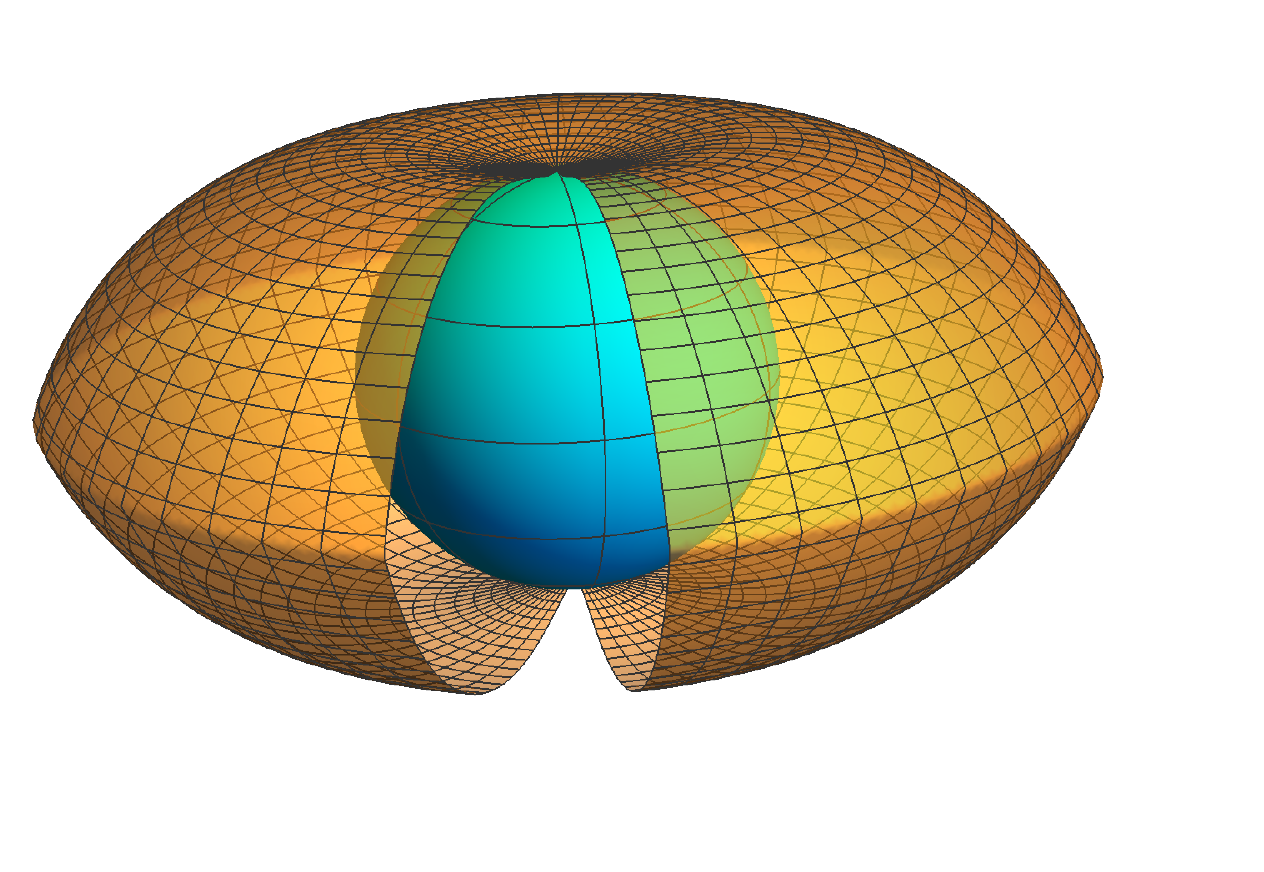}
}}
\centerline{(a)~~~~~~~~~~~~~~~~~~~~~~~~~~~~~~~~~~~~~~~~~~~~~~~~~~~~~~~(b)}
\caption{\textbf{Polar plots of the excitation energy gaps.} (a) The chiral $d$-wave singlet case and (b) the nonunitary chiral $p$-wave triplet case. In both the cases, point nodes appear at the two ``poles'', while the singlet case has an additional line node at the ``equator''.}
\mylabel{fig:gap}
\end{figure*}

The strong SOC case considered here implies that the single particle states are no longer the eigenstates of spin and we need to label them rather by pseudospins. The pseudospin states are linear combinations of the spin eigenstates. Since the pseudospin and the spin are closely related, the even parity states correspond to pseudospin singlet and the odd parity states correspond to pseudospin triplet states. 

We can now follow the standard Bogoliubov-de Gennes mean field theory~\cite{Sigrist1991} to compute the quasi-particle excitation energy spectrum for the two TRS breaking states given in \eqn{eqn:singlet} and \eqn{eqn:triplet}. The schematic view of the excitation energy gaps for the two order parameters are shown in \fig{fig:gap}.

\subsection{Topological properties of the chiral singlet state}
To discuss the topological properties of the nodal excitations for the chiral $d$-wave state with the gap function
\beq
\Delta(\bk) = \frac{\Delta_0}{k^2_F}k_z(k_x + i k_y)
\eeq
with $\Delta_0$ being the pairing amplitude, we assume a simplified single band parabolic dispersion (in units of $\hbar$)
\beq
\xi(\bk) = \frac{k^2}{2m} - \mu,
\eeq
where $m$ is the mass of an electron, $\mu =\frac{k^2_F}{2m} $ is the chemical potential and $k_F$ is the Fermi wavevector. We note that $\Delta(\bk) \sim Y^2_1(\theta,\phi)$ where $Y^l_m(\theta,\phi)$ are the spherical harmonics. Thus the Cooper pairs have an angular momentum $L_z = +1$ for this state.

 Then Bogoliubov-de Gennes Hamiltonian in the pseudospin basis can be written as
\beq
\mathcal{H} = \sum_{\bk} \Psi^{\dagger}_{\bk} H(\bk) \Psi_{\bk}
\eeq
where $\Psi_{\bk} = (c_{\bk\uparrow}, c^{\dagger}_{-\bk\uparrow})^T$ with $c_{\bk\sigma}$ being the fermion annihilation operator with pseudospin flavor $\sigma \in \{\uparrow,\downarrow\}$. We can rewrite the BdG Hamiltonian as
\beq\mylabel{eqn:chiral_Ham}
H(\bk) = \mathbf{N}(\bk)\cdot\pmb{\tau}
\eeq   
where $\pmb{\tau}$ is the vector of the three Pauli matrices in the particle-hole space and $\mathbf{N}(\bk) = \left\{\frac{\Delta_0}{k^2_F} k_z k_x, \frac{\Delta_0}{k^2_F}k_z k_y, \xi(\bk) \right\}$ is a pseudospin vector. The eigenvalues of the Hamiltonian in \eqn{eqn:chiral_Ham} are $\pm E(\bk)$ where
\beq
E(\bk) = |\mathbf{N}(\bk)| = \sqrt{\xi^2(\bk) + |\Delta(\bk)|^2}.
\eeq
Hence, the superconducting ground state has two point nodes at the two poles of the Fermi surface $\bk_\pm=(0,0,\pm k_F)$ and a line node at the equator $k_z=0$ plane. The low energy Hamiltonian close to two point nodes can be written as 
\beq
H(\bk) = \frac{\Delta_0}{k_F}(p_x \tau_x -p_y \tau_y) \pm v_F p_z \tau_z
\eeq
where we have defined $\bp = (\bk - \bk_\pm)$. This is a Weyl Hamiltonian. Thus the two point nodes are also Weyl nodes. As a result they are impossible to gap out since there is no fourth Pauli matrix which can come from a mass term to gap out the nodes.

The corresponding Bloch wave functions $\ket{u_{\pm}(\bk)}$ are the eigenfunctions of $\mathbf{\hat{n}}(\bk).\pmb{\sigma}$ with eigenvalues $\pm 1$ where $\mathbf{\hat{n}}(\bk) = \mathbf{N}(\bk)/|\mathbf{N}(\bk)|$ is the unit vector along the direction of the pseudospin $\mathbf{N}(\bk)$. We note that this unit vector $\mathbf{\hat{n}}(\bk)$ is well defined only when $|\mathbf{N}(\bk)| \neq 0$ i.e. in the nodeless regions on the Fermi surface. In spherical coordinates, parametrizing $\mathbf{\hat{n}}(\bk) = [n_x(\bk), n_y(\bk), n_z(\bk)] = [\sin(\theta)\cos(\phi), \sin(\theta)\sin(\phi),\cos(\theta)]$ we have
\beq
\ket{u_{-}(\bk)}=\begin{bmatrix}
\cos(\frac{\theta}{2}) e^{-i\phi} \\ \sin(\frac{\theta}{2})
\end{bmatrix} \,\,\text{and}\,\, \ket{u_{+}(\bk)}=\begin{bmatrix}
\sin(\frac{\theta}{2}) e^{-i\phi} \\ -\cos(\frac{\theta}{2})
\end{bmatrix}.
\eeq   
Then from the negative energy occupied states $\ket{u_{-}(\bk)}$ the Berry connection is defined as
\beq
\mathbf{A}(\bk) = i \bra{u_{-}(\bk)}\pmb{\nabla}_{\bk}\ket{u_{-}(\bk)}
\eeq
and the corresponding Berry curvature is $\mathbf{F}(\bk) = \pmb{\nabla}_{\bk} \times \mathbf{A}(\bk)$. In terms of the components of $\mathbf{\hat{n}}(\bk)$, it is given by $\mathbf{F}(\bk) = [n_y(\bk)\{\pmb{\nabla}_{\bk}n_z(\bk) \times \pmb{\nabla}_{\bk}n_x(\bk)\} - n_x(\bk)\{\pmb{\nabla}_{\bk}n_z(\bk) \times \pmb{\nabla}_{\bk}n_y(\bk)\}]/[2\{n^2_x(\bk)+n^2_y(\bk)\}]$.

For the chiral $d$-wave case, $F_x(\bk)$ and $F_y(\bk)$ are odd functions of $(k_y, k_z)$ and $(k_x, k_z)$ respectively. Hence, there is no Berry flux along the $x$ and $y$ directions. The number of field lines coming in and out of the $ca$ and $cb$ planes are the same. Whereas $F_z(\bk)$ is an even function of $(k_x, k_y)$ and the flux through the $ab$ plane as a function of $k_z$ is 
\beq
\Phi(\bk) = \int dk_x dk_y F_z(\bk) = 2\pi \mathcal{C}(k_z).
\eeq
$\mathcal{C}(k_z)$ is the "sliced" Chern number (momentum dependent) of the effective $2$D problem for a fixed $k_z$. For a given value of $|k_z|<k_F$, the Hamiltonian in \eqn{eqn:chiral_Ham} describes an effective $2$D problem with fully gapped weak coupling BCS pairing and an effective chemical potential $\frac{\hbar^2}{2m}(k^2_F - k^2_z)$ having the Chern number $C(k_z) = +1$. For $|k_z|>k_F$, the effective chemical potential is negative and describes a topologically trivial BEC state. Thus, the Weyl point nodes at $(0,0,\pm k_F)$ act as monopoles and anti-monopoles of the Berry curvature and the flux through a sphere surrounding the monopole is $2\pi$ and that through the anti-monopole is $-2\pi$. The topologically protected Weyl nodes give rise to Majorana arc surface states on the surface Brillouin zone corresponding to the $(1,0,0)$ and $(0,1,0)$ surfaces having chiral linear dispersions along $y$ and $x$ directions respectively. As a result of the arc surface states the system shows anomalous thermal and spin Hall effects~\cite{Schnyder2015,Goswami2015,Goswami2013}.

The equatorial line node is characterized by a 1D winding number. This can be defined in terms of the following spectral symmetry~\cite{Goswami2015,Goswami2013} of the Hamiltonian. We note that the operator 
\beq
\Gamma_{\bk} = \sin(\phi_{\bk}) \tau_x + \cos(\phi_{\bk})\tau_y
\eeq
where $\tan(\phi_{\bk}) = k_y/k_x$ anticommutes with the Hamiltonian
\beq
\{H(\bk),\Gamma_{\bk}\} = 0.
\eeq 
As a result any eigenstate of the Hamiltonian $H(\bk)$ with the eigenvalue $E_{\bk}$ is also an eigenstate of the operator $\Gamma_{\bk}$ with the eigenvalue $-E_{\bk}$. Then with the help of this spectral symmetry $\Gamma_{\bk}$ we define the winding number as
\beq
w({\bk}_{\perp}) = - \frac{1}{4\pi i} \oint\limits_{\mathcal{L}}\,\mathrm{d}l\,\, Tr\left[ \Gamma_{\bk} H^{-1}(\bk)\partial_l H(\bk)\right], 
\eeq
where $dl$ is the line element along a closed loop $\mathcal{L}$ encircling
the line node and ${\bk}_{\perp}=(k_x,k_y)$. For this case then we have
\bea
w({\bk}_{\perp}) &=& 1 \,\,\,\,\forall\, k_{\perp} < k_F\\
&=& 0 \,\,\,\, {\rm otherwise}.
\eea
We note that the winding number does not depend on the angular momentum of the Cooper pairs. This nontrivial topology of the line node ensures the existence
of zero-energy surface Andreev bound states on the $(0,0,1)$ surface. They produce an image of the Fermi surface equator in the corresponding surface Brillouin zone. Being dispersionless, these zero-energy states result in a divergent density of states, and are predicted to give rise to a zero
bias peak in tunneling measurements. These zero modes are two fold degenerate Majorana fermions arising from the twofold spin degeneracy of the pairing interaction.

\bibliography{chiral.bib}

\end{document}